\documentclass{emulateapj}
\shorttitle{Detection of the Transverse Proximity Effect}
\slugcomment{Received 2007 August 16; Accepted 2007 November 20}

\newcommand{\beq}{\begin{equation}}
\newcommand{\eeq}{\end{equation}}
\newcommand{\ee}[1]{\times 10^{#1}}
\newcommand{\units}[1]{\;{\rm #1}}

\newcommand{\cc}{\c{c}}

\newcommand{\lya}{\ensuremath{\rm Ly\alpha}}

\newcommand{\kms}{\rm km~s\ensuremath{^{-1}\,}}

\newcommand{\novi}{$N$(O\,{\sc vi})}
\newcommand{\nciv}{$N$(C\,{\sc iv})}
\newcommand{\nhi}{$N$(H\,{\sc i})}
\newcommand{\ovi}{O\,{\sc vi}}
\newcommand{\civ}{C\,{\sc iv}}
\newcommand{\hi}{H\,{\sc i}}

\newcommand{\secpoint}{\mbox{$''\mskip-7.6mu.\,$}}

\def\ltsima{$\; \buildrel < \over \sim \;$}
\def\simlt{\lower.5ex\hbox{\ltsima}}
\def\gtsima{$\; \buildrel > \over \sim \;$}
\def\simgt{\lower.5ex\hbox{\gtsima}}
\def\arcs{$''~$}
\def\arcm{$'~$}

\usepackage{amssymb}
\usepackage{epstopdf}
\usepackage{graphicx}

\def\alwaysmath#1{\ifmmode{#1}\else{$#1$}\fi}
 
\shortauthors{Gon\cc alves, Steidel, \& Pettini}

\begin{document}

\title{DETECTION OF THE TRANSVERSE PROXIMITY EFFECT: RADIATIVE FEEDBACK FROM BRIGHT
QSOs\altaffilmark{1}}
\author{\sc Thiago S. Gon\cc alves and Charles C. Steidel}
\affil{California Institute of Technology, Mail Stop 105-24, Pasadena, CA 91125}
\author{\sc Max Pettini}
\affil{Institute of Astronomy, Madingley Road, Cambridge CB3 OHA, UK}

\altaffiltext{1}{Based on data obtained at the W.M. Keck Observatory, which is operated
as a scientific partnership among the California Institute of Technology, the University
of California, and NASA, and was made possible by the generous financial support of the W.M.
Keck Foundation.}


\begin{abstract}

Measuring the response of the intergalactic medium to a blast of ionizing radiation allows one to infer the
physical properties of the medium and, in principle, the lifetime and isotropy of the radiating source. The most
sensitive such measurements can be made if the source of radiation is near the line of sight to a bright
background QSO. We present results based on deep Keck/HIRES observations of the QSO triplet KP76, KP77 and KP78
at $z \simeq 2.5$, with separations of 2-3\arcm\ on the plane of the sky. Using accurate systemic redshifts
of the QSOs from near-IR spectroscopy, we quantify the state of the IGM gas 
in the proximity regions where the expected 
ionizing flux from the foreground QSOs exceeds that of the metagalactic
background by factors $10-200$, assuming constant and isotropic emission. Based on the unusual ionization
properties of the absorption systems with detected \hi, \civ\ and \ovi, we conclude that the gas has been
significantly affected by the UV radiation from the nearby QSOs. Aided by observations of the galaxy
density near the foreground QSOs, we discuss several effects that may explain why the transverse
proximity effect has eluded most previous attempts to detect it. 
Our observations suggest that the luminosities of KP76 and KP77 have
remained comparable to current values over timescales of, respectively, 
$\Delta t > 25$ Myr and 16 Myr $< \Delta t
< 33$ Myr - consistent with typical QSO lifetimes estimated from independent, less-direct methods. 
There is no evidence that the UV radiation from either QSO was significantly anisotropic during these
intervals. 

\end{abstract}


\section{INTRODUCTION}

The QSO ``proximity effect'' refers to the observation that 
the incidence of Lyman $\alpha$ (Ly$\alpha$)
absorption from the IGM
decreases at redshifts close to that of the QSO.
If this deficit is interpreted as being due to the influence of 
the radiation field of the QSO on the nearby IGM, 
then measurements of the \hi\ density near QSOs of known 
luminosity may be used to determine the intensity of the 
metagalactic ionizing background \citep{murdoch86,bajtlik88,scott00}.
The statistical significance of the QSO proximity effect 
is high enough that there is little doubt of its reality.
However, the measurement of the intensity of the 
metagalactic radiation field 
from observations of the Ly$\alpha$ forest 
is subject to a number of systematic uncertainties, 
in particular the uncertainty in the QSO systemic redshift 
(e.g., \citealt{scott00,espey89}), 
the possibly over-dense large-scale environment
in which QSOs may be located (e.g., \citealt{faucher07}), 
and the non-linear response of the observed
H\,{\sc i} optical depth to changes in the radiation field 
intensity, such that only \hi\ systems with
relatively low \hi\ optical depths 
($\tau_{\rm H\,I} \simlt 10$) will be significantly altered. 
All of these effects would lead to an overestimate of the 
background intensity through an underestimate of the 
sphere of influence of radiation from the QSO: 
QSO redshifts determined from the broad UV emission
lines tend to be lower than the true systemic redshifts, 
so that the true ``proximity region'' where the QSO radiation field
dominates over the background is actually larger than inferred.
Similarly, if QSOs are preferentially found in
overdense regions, then the incidence of \hi\ 
with sufficient optical depth to remain relatively
unscathed by an enhanced radiation field will be 
higher than at an average location in the universe,
suggesting that the QSO's effect on the local radiation field 
is smaller than it really is. 
Clearly, there are also statistical (as opposed to systematic) 
issues associated with the proximity effect, 
since one expects relatively large sample variance 
in the \hi\ content between the relatively small
volumes probed by individual QSOs.

The so-called ``transverse proximity effect'' (hereinafter, TPE), 
on the other hand, involves searching for
the influence of foreground sources of UV radiation 
on the IGM absorption observed in the spectra
of background objects. 
Unlike the line-of-sight proximity effect, 
measurements of the transverse
proximity effect can in principle constrain the 
radiative lifetime {\it and} the isotropy
of the sources (e.g., \citealt{schirber04}). 
Line of sight measurements of the proximity effect 
are sensitive only to fluctuations in the source intensity 
on timescales of $\simlt 10^4$ yr (e.g., \citealt{martini04}),
whereas transverse measurements can sample longer 
timescales---determined by the light travel time 
between the foreground source
and the observed line of sight---that 
are more interesting in the context 
of QSO lifetimes. 
The sphere of influence of the radiation field from a source
of UV photons, as for the line of sight proximity effect, 
depends on luminosity of the source.
For a typical bright QSO ($V \sim 18$) at $z \sim 2.5$, 
the sphere within which an isotropically
radiating QSO significantly dominates 
over the metagalactic background has
a physical radius of $r \sim 5-10$\,Mpc,
corresponding to a line of sight velocity 
range of $\Delta v_{\rm r} \simeq 1250-2500$\,\kms\
for material moving with the Hubble expansion.
The corresponding light travel time is $15-30$\,Myr. 
A number of quasi-independent arguments have suggested 
that bright QSO typically have lifetimes of $10^6-10^8$\,yr (see, e.g.,
\citealt{haehnelt98,richstone98,martini01,hosokawa02,yu02,steidel02}) 
so that the potential sphere of influence of the UV radiation
from bright QSOs and the timescale of QSO `events' 
may fortuitously  be of the same order. 
The relevant
angular scales are $\theta \simlt 10$\arcm\ 
on the sky at $z \sim 2-3$, with the amplitude of the expected effects
varying as $1/\theta^2$ and 
proportional to the (far-UV) luminosity of the sources.  

A number of authors have used QSOs 
with small angular separations on the sky, 
but at different redshifts, to search for the TPE 
via observations of the Ly$\alpha$ forest
(e.g., \citealt{fern-soto95,crotts98,croft04,schirber04}). 
To date, as far as we are aware, no evidence
for a reduction in the number of Ly$\alpha$ 
forest lines at redshifts near foreground QSOs
has been found\footnote{\citet{gallerani07} observe
a ``transmission gap'' in the Lyman $\alpha$ forest of
a background $z=6.42$ QSO near the redshift of a foreground QSO at $z=5.65$, although
the statistical significance of the region of reduced Ly $\alpha$ opacity is difficult
to evaluate.}. 
In fact, all of the above-referenced works 
except \citet{fern-soto95} reported instead
an {\it excess} of Ly$\alpha$ absorption systems 
at the redshifts where TPE deficits would have
been expected. 
The common explanations for the lack of the expected signal 
include some combination of the QSO large-scale environment, 
short or intermittent radiative lifetimes of
the QSOs, and anisotropy of the QSO radiation. 
More recently, \citet{hennawi06} found that
high column density H\,{\sc i} systems at the redshifts 
of foreground QSOs are much more common than   
associated systems of similarly high column density
seen in the spectra of QSOs at their own redshifts.
These authors also estimated that 
the clustering of the high column density \hi\ 
systems with the QSOs may be even stronger
than that of star-forming galaxies at the same redshifts. 
Such observations have been interpreted as evidence that
H\,{\sc i} gas is distributed anisotropically 
around QSOs, possibly because of QSO variability
and/or anisotropic UV radiation.  
As a counter-argument, however, observations of
the He\,{\sc ii}, rather than H\,{\sc i}, Ly$\alpha$ forest
seem to show that foreground QSOs do in fact 
have a measurable effect on the ionization state of gas 
within volumes corresponding to light travel times of 
$\sim 10^6-10^7$ Myr \citep{jakobsen03,worseck06,worseck07}. 
This may suggest that H\,{\sc i} is not the most sensitive barometer
of changes in the radiation field near QSOs.  
The key advantage of the He\,{\sc ii} transition is that it is sensitive to the intensity
of high-energy photons that can plausibly only be produced by QSOs or AGN, so that one can
sense changes in both the intensity and the shape of the radiation field for local sources. 

A phenomenon closely related to the TPE 
is the fluorescence of high column density \hi\
gas in the vicinity of bright sources of ionizing photons 
(see, e.g., \citealt{gould96,cantalupo05}). 
Searches for fluorescence at the redshifts of QSOs 
have so far yielded mixed results. 
\citet{adelberger06} identified what appears to be a 
damped Ly$\alpha$ system (DLA) at $z=2.84$
that is producing Ly$\alpha$ emission in response 
to the ionizing radiation from a very bright QSO 
located $\sim 380$\,kpc away, implying that
the QSO luminosity has remained approximately constant
over the last $\sim 10^6$\,yrs (and that the QSO is
radiating isotropically).
However, in another case,
the lack of fluorescence around a QSO at a similar redshift
was interpreted by \citet{francis04}
as evidence for anisotropic emission from the QSO.

Thus, the observational situation 
concerning the TPE is currently ambiguous. 
QSO lifetimes of $>10^6$\,yrs and 
largely isotropic QSO radiation are supported by
some observations, but challenged by others. 
For this reason, we have elected to investigate the TPE 
using a different approach from that of most previous attempts. 
Rather than counting Ly$\alpha$ lines or evaluating
the average Ly$\alpha$ opacity near foreground QSOs, 
we focus on the metal line systems.
Our strategy is to closely examine the
ionization state of metal line systems 
within the expected spheres of influence of
foreground QSOs, using the combination of
very high quality optical spectra of the background QSOs
and precise determinations of the systemic
redshifts of the foreground QSOs
from near-infrared (near-IR) spectroscopy.
Our aim is to determine the strength of 
the ionizing radiation field seen by gas 
at various distances from the foreground QSOs, 
independently of the statistical limitations and 
systematic uncertainties inherent in line counting 
or ``mean flux'' techniques. 
The hope is that the new data will 
help to explain why 
previous results have been so mixed,
as well as constrain better the 
radiative lifetime and solid angle 
of QSO emission.

\begin{figure}[htb]
\plotone{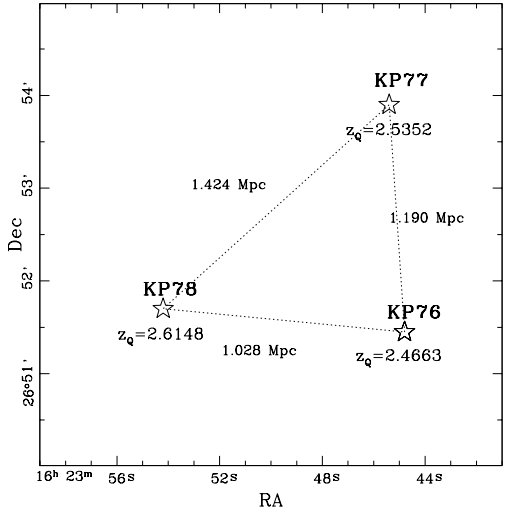}
\caption{The Q1623-KP76, KP77, KP78 triplet on the plane of the sky. 
The transverse physical distances between
the QSO sightlines (evaluated at the redshift of 
the lower redshift QSO in each case) are indicated.  
The QSO systemic redshifts were 
measured from accurate near-IR spectroscopy (\S 2.2). }
\label{triplet}
\end{figure}

\begin{figure*}[htb]
\plotone{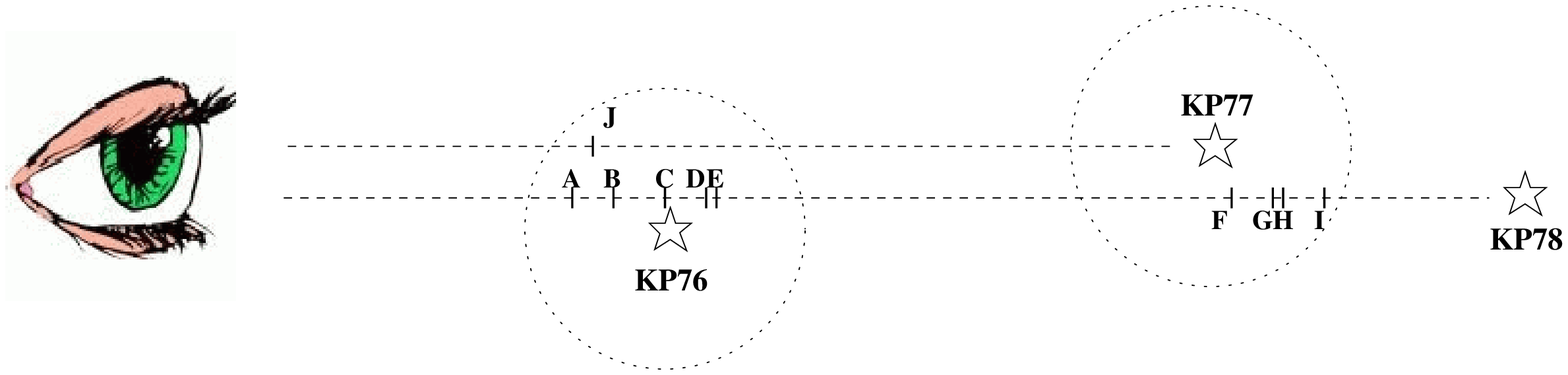}
\caption{Schematic view of the QSO triplet.
The relative positions
of the metal line absorption systems studied in this paper,
labeled A--J,  
are indicated within dotted circles representing the
proximity regions of the foreground QSOs KP76 and KP77. 
Within these volumes---spheres of physical radius
$\sim 5$\,Mpc--- the ionizing flux from the nearby
QSO is expected to exceed that due to the metagalactic
background by more than one order of magnitude,
if the QSOs radiate isotropically and their luminosities
have remained approximately 
constant over the last 20--30\,Myr.
}
\label{scheme}
\end{figure*}

In the first stage of this program reported here,
we focus on the QSO triplet KP76, KP77, KP78 
(also known as Q1623+268).
This grouping is among the best studied in the literature,
having figured prominently in earlier work aimed 
at measuring the coherence scale
and clustering of the Ly$\alpha$ forest (e.g., \citealt{sargent82,crotts89,crotts98,crotts97}).
The relative configuration of the three QSOs is 
illustrated in Figures~\ref{triplet} and \ref{scheme}. 
\citet{crotts98} searched for the TPE 
in the Ly$\alpha$ forests of this triplet
and concluded---based on counts of Ly$\alpha$ absorption 
lines---that there is no evidence that the foreground QSOs
(KP76 and KP77) reduce the density of Ly$\alpha$ lines
in the expected proximity regions. 
Rather,  the line density appears to be higher 
than at an average location in the forest at
comparable redshifts.

In this paper, we present the highest quality 
high resolution spectra of these QSOs to date, 
obtained with the recently upgraded 
High Resolution Echelle Spectrograph (HIRES)
on the Keck~I telescope. We have also obtained, as
part of an ongoing survey of galaxies and the IGM 
at redshifts $1.8 \simlt z \simlt 3$ (see \citealt{steidel04,
adelberger05}), spectra of approximately 300 other galaxies, 
QSOs, and AGN in this redshift range
within a $\sim 15$\arcm\ field centered on the QSO triplet, 
providing unprecedented information on the surrounding 
large scale structure.  
We examine the detailed properties of 
absorption line systems within the expected proximity
regions of the two foreground QSOs, and we interpret 
our findings in the context of our knowledge
of the distribution of galaxies in the same regions. 

\section{OBSERVATIONS}
\label{sec:observations}

\subsection{HIRES Spectroscopy}

The Q1623+2651 field \citep{sramek78}, which includes
the QSO triplet KP76, KP77 and KP78, is part of
a large-scale survey of galaxies and 
AGN in the fields of bright QSOs that we have been 
conducting during the past several years \citep{steidel04}. 
The three QSOs have $u'$ band magnitudes (measured from our photometry) 
of 18.62 (KP76; $z_{\rm em} =2.4663$), 
17.48 (KP77; $z_{\rm em} = 2.5352$), and 18.82 
(KP78; $z_{\rm em} = 2.6148$), 
presenting a very unusual configuration of
three bright QSOs at similar redshifts and within a few arcmin on the sky. 
The angular separations between pairs are 127\arcs\
(KP76--KP78), 147\arcs\ (KP76--KP77) and 177\arcs\ (KP77--KP78), 
corresponding to proper transverse distances
(evaluated at the redshift of the lower redshift member of each pair) 
of 1.028, 1.190, and 1.424 physical Mpc respectively, 
for the $\Omega_{\rm M}=0.3$, 
$\Omega_{\Lambda}=0.7$, $H_0 = 70$\,km~s$^{-1}$~Mpc$^{-1}$
($h=0.7$) cosmology that is assumed throughout this paper
(see Figure~\ref{triplet}).

The observations reported here were obtained 
on the nights of 2005 May 31,  June 1, and October 9 and 10, 
with the Keck~I telescope and HIRES \citep{vogt94} using
the UV cross-disperser and UV sensitive CCD array. 
The data cover the wavelength range $3100-6000$\,\AA, 
with small gaps in spectral coverage near 4000\,\AA\
and 5000\,\AA\ corresponding to gaps between CCD chips  
on the detector. 
All exposures were taken 
through the 1\secpoint148 slit, 
resulting in a resolution of
8.5\,\kms (FWHM), sampled with 
$\sim 3$ pixels per resolution element. 
Some additional HIRES spectra of KP76 and KP77, 
obtained in 1995 August by W. Sargent and
collaborators, were also included. 
The {\sc makee} data reduction package 
written by Tom Barlow was used to process
the two-dimensional HIRES spectra, extract
them and map them onto   
a vacuum heliocentric wavelength scale,
and finally combine them; continuum
fitting prior to the merging of echelle orders 
was accomplished using the {\sc imanip} 
package, kindly provided by Robert Simcoe.  
The final one-dimensional spectra have 
typical signal-to-noise (S/N) ratios of $15-40$ per 2.8\,\kms pixel. 

\subsection{NIRSPEC Spectroscopy}

Given the importance of establishing the systemic 
redshifts of the QSOs, in order to accurately evaluate their effect
on the nearby IGM, we obtained near-IR
spectra of all three QSOs encompassing the 
[O\,{\sc iii}]\,$\lambda\lambda 4959$, 5007 
and H$\beta$ emission lines.
The spectra were obtained in 2004 September with 
the Near Infrared Spectrograph (NIRSPEC; McLean et al. 1998)
on the Keck~II telescope in the low resolution 
($R \simeq 1400$) mode using the NIRSPEC-5 filter
(corresponding approximately to the near-IR $H$ band). 
The reductions were performed in the manner 
described in detail by \citet{erb06}. 

The redshifts of KP76 and KP77 were 
determined from Gaussian fits to the [O\,{\sc iii}] emission lines 
yielding $z = 2.4663 \pm 0.0003$ and $z = 2.5353 \pm 0.0003$, 
respectively. The redshift of KP78, which is less
crucial to our analysis since this QSO 
lies in the background of the other two,
was measured to be $z=2.6148 \pm 0.0004$ 
from the H$\beta$ line because 
[O\,{\sc iii}] falls in a region of very poor 
atmospheric transmission between the $H$ and $K$ bands. 
Note that published catalog redshifts
for the three QSOs are 2.490, 2.5177, and 2.6017, respectively 
(the last two from the third data release of the Sloan Digital
Sky Survey); they differ from the newly determined, and more
accurate, systemic redshifts by  $+2050$, $-1490$, and $-1090$ \kms
respectively.     
In our work, we have found that such large redshift offsets
are in fact typical of bright QSOs, and it is worthwhile 
pointing out that they are 
as large as, or larger than, the velocity range 
corresponding (for material moving with the Hubble flow) 
to distances
over which the radiation field of a bright QSO 
dominates over the background (see \S~3 below).
Thus, it is essential to determine accurate
redshifts from near-IR spectroscopy for a meaningful
assessment of the proximity effect.

Even so, the systematic uncertainties in the redshifts of the three
QSOs are likely to exceed the statistical error of $\pm 30$\,\kms 
of our Gaussian fits.  
For example, \citet{boroson05} has shown, using
a large sample of low redshift AGN, 
that the redshifts defined by the [O\,{\sc iii}] doublet
are lower than those of the host galaxies by
40\,\kms on average, and that the offset can be up to 
ten times higher in the most extreme cases. 
Nevertheless, the [O\,{\sc iii}] doublet is
a far better indicator of systemic redshift 
than the broad UV emission lines commonly used.
For the purposes of the present work, we 
will assume that the above values are the 
systemic redshifts of the three QSOs considered here.  

\section{IDENTIFICATION OF ABSORBERS AND ANALYSIS}
\label{sec:absorbers}

\begin{figure*}[htb]
\plotone{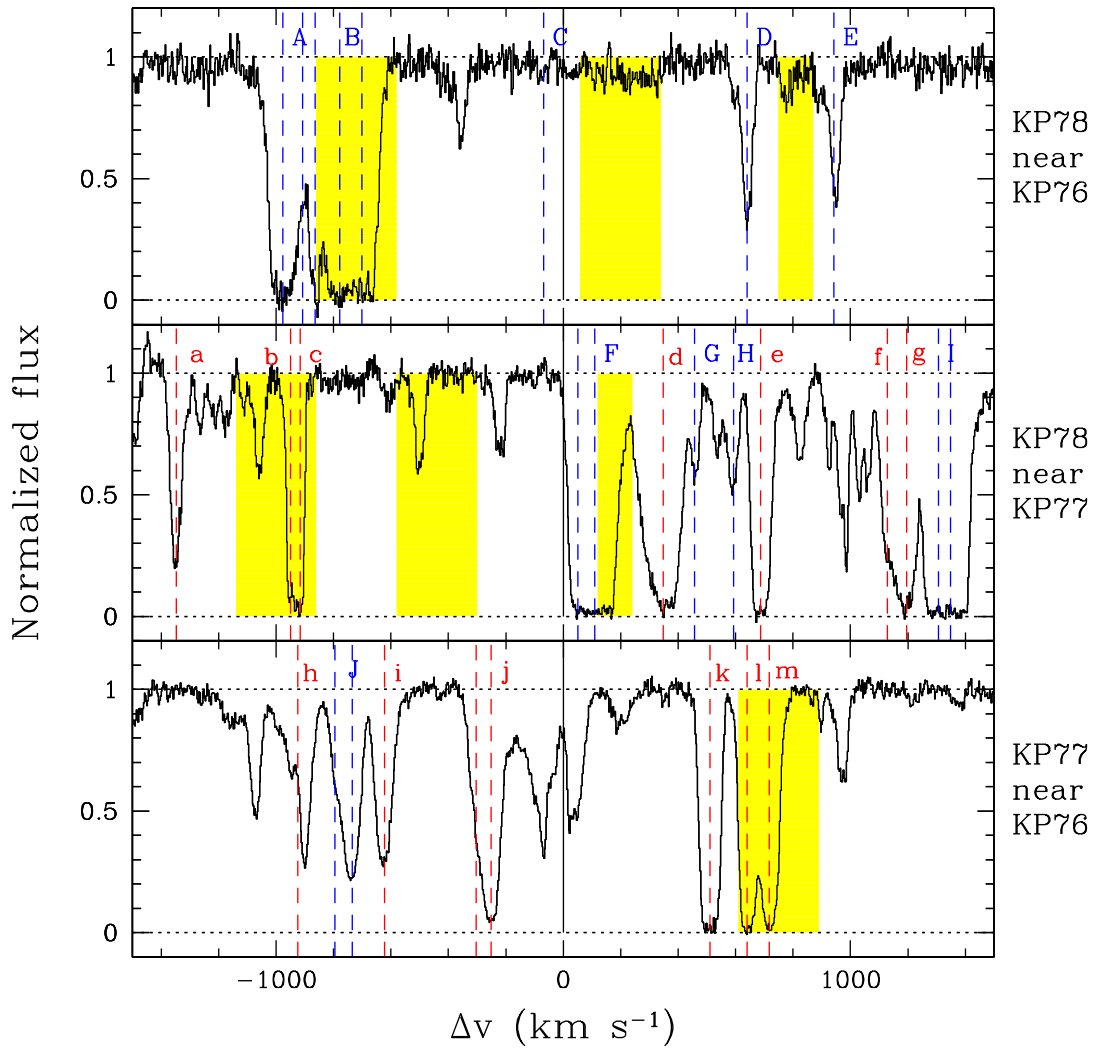}
\caption{Portions of the spectra of the background 
QSOs near the redshifts of foreground QSOs. 
Each panel shows the Ly$\alpha$  forest within 
$\pm 1500$\,km~s$^{-1}$ 
of the foreground QSO redshift.  
The ten metal line systems identified within these proximity regions 
are denoted by letters A--J  and by blue dashed lines, while
13 \hi-only absorbers with $\log N$(\hi)$> 13.5$ 
are indicated by lower-case letters and red dashed lines.  
The yellow shaded regions mark the redshifts (together 
with their uncertainties) of seven spectroscopically confirmed
galaxies within the proximity regions of KP76 and KP77 and
close to the sightlines to KP77 and KP78. 
These galaxies are discussed in \S\ref{subsec:gals}.
}
\label{deltaz}
\end{figure*}

The first step in the analysis is to determine the 
wavelength intervals over which to search for evidence 
of the TPE in the HIRES spectra of the two
background QSOs, KP78 and KP77. 
These intervals correspond to the physical distances to
which the ionizing flux from the foreground QSOs 
will significantly dominate over the metagalactic radiation field. 
Given the difficulties experienced by previous observers in detecting
the TPE, we decided to be deliberately conservative in the choice of interval 
and limited our analysis to 
the volume within which the expected 
radiation field intensity from the proximate QSOs exceeds the 
metagalactic radiation field intensity by a factor $\simgt 10$, where
the impact on the IGM should be easiest to recognize. 
We calculated the QSO luminosities using their measured broadband u$^{\prime}$ fluxes
(corresponding to AB magnitudes at rest-frame wavelengths of $\simeq 1000$\AA, just above the Lyman
limit at 912\AA); these values were corrected for line blanketing in the Lyman $\alpha$ forest
estimated from the spectra themselves, increasing their brightness by $\simeq 20$\%.  
We adopted a metagalactic background intensity at the Lyman limit of
$J_{\rm bg}(\nu_0) \simeq 5 \times 10^{-22}$\,ergs~s$^{-1}$~cm$^{-2}$~Hz$^{-1}$~sr$^{-1}$ (Scott et al. 2000; Tytler et al. 2004; Bolton et al. 2005).
Using these numbers, the ionizing fluxes from KP76 and KP77 will
exceed that from the metagalactic radiation field 
by factors of $\simgt 10$
within spheres of physical radius $\sim 5$\,Mpc,
assuming that the 
QSOs radiate isotropically and that their luminosities have
remained approximately constant over the relevant transverse 
light travel times of $\sim 10^6-10^7$ yrs.
In our cosmology, a radius of 5\,Mpc corresponds to a
velocity interval $\Delta v \simeq \pm 1250$\,\kms
for material moving with the Hubble expansion.
Considering the uncertainty in the systemic redshifts
of the QSOs which, from the discussion above we estimate 
to be $\pm 100$\,km~s$^{-1}$, and an additional uncertainty
of $\pm 200$\,km~s$^{-1}$ to account for departures 
from the Hubble flow due to the local density field,
we settled on a velocity range 
$\Delta v \simeq \pm 1500$\,km~s$^{-1}$,
centered on the systemic redshifts of 
KP76 and KP77, over which
to examine the properties of the IGM in the HIRES
spectra of KP77 and KP78. 
These ``proximity regions'' in the relevant portions of
the Ly$\alpha$ forest are shown in Figure~\ref{deltaz}.

We identified metal line systems within these proximity regions
by the presence of Ly$\alpha$ absorption and either
C\,{\sc iv}\,$\lambda\lambda 1548,1550$ or
O\,{\sc vi}\,$\lambda\lambda 1031,1036$ (or both).
Since the first-order effect
of an enhanced ionizing radiation field 
is to reduce the \hi\ optical depth $\tau_{\rm H\,I}$, 
we did not require that strong
\hi\ be present; in fact, one system (System C below) 
has very weak \hi\  
that would not have been identified had
the search been limited by Ly$\alpha$ optical depth.   
The wavelength coverage of the HIRES 
spectra down to 3100\,\AA\ in the observed frame
is important for confirming that a given absorption
feature in the forest is  indeed Ly$\alpha$ (based
on the presence of Ly$\beta$) and for 
accurate measurement of the \hi\ column density
$N$(H\,{\sc i})
in cases where the Ly$\alpha$
line is saturated (using higher order Lyman lines).  
Many of the Ly$\alpha$ lines within the 
proximity regions have no associated 
C\,{\sc iv} or O\,{\sc vi} absorption even at the 
sensitive detection limits reached by our HIRES spectra.
In the subsequent analysis, 
we include all such H\,{\sc i}-only systems
with column densities \nhi$ > 10^{13.5}$~cm$^{-2}$,
for which we have confidence in both their 
identification and their measured
column densities through 
the detection of higher order Lyman lines.
In general these H\,{\sc i}-only systems
do not provide significant constraints on the ionization level
of the IGM, but they are included for completeness. 
The $3 \sigma$ upper limits to metal line column densities 
are typically $\log N$(O\,{\sc vi})\,$ \leq 12.7$ and
$\log N$(C\,{\sc iv})\,$ \leq 12.1$
(with $N$ in units of cm$^{-2}$).

\begin{deluxetable}{lcccc}
\tablecaption{Metal-Line Systems in Proximity Regions}
\tablecolumns{5}
\tablewidth{0pt}
\tabletypesize{\scriptsize}
\tablehead{
\colhead{System} & \colhead{Ion}  & \colhead{$z$}  & \colhead{$\log N$ (cm$^{-2}$)}  & \colhead{$b$ (km s$^{-1}$)}
}
\startdata
System A & \hi\  & 2.4550 & 14.76 $\pm$ 0.02 & 31.4 $\pm$ 0.8\\
& \hi\  & 2.4558 & 13.36 $\pm$ 0.13 & 20.3 $\pm$ 6.3\\
& \ovi\ & 2.4557 & 13.87 $\pm$ 0.14 & 41.8 $\pm$ 15.0\\
& \civ\ & 2.4550 & 12.92 $\pm$ 0.04 & 18.9 $\pm$ 1.9\\
\\
System B & \hi\  & 2.4563 & 14.12 $\pm$ 0.03 & 17.5 $\pm$ 1.3\\
& \hi\  & 2.4573 & 14.93 $\pm$ 0.04 & 33.2 $\pm$ 1.8\\
& \hi\  & 2.4582 & 14.83 $\pm$ 0.03 & 31.8 $\pm$ 1.2\\
& \ovi\ & 2.4578 & 14.12 $\pm$ 0.06 & 28.5 $\pm$ 4.5\\
& \ovi\ & 2.4584 & 14.03 $\pm$ 0.07 & 20.5 $\pm$ 3.4\\
& \civ\ & 2.4564 & 12.87 $\pm$ 0.04 & 9.5  $\pm$ 1.3\\
& \civ\ & 2.4578 & 12.68 $\pm$ 0.09 & 8.5  $\pm$ 2.9\\
& \civ\ & 2.4583 & 12.85 $\pm$ 0.07 & 13.5 $\pm$ 2.8\\
\\
System C & \hi\  & 2.4654 & 12.44 $\pm$ 0.07 & 23.9 $\pm$ 4.8\\
& \ovi\ & 2.4656 & 14.62 $\pm$ 0.02 & 34.2 $\pm$ 0.9\\
& \civ\ & ... & $\leq 12.13$ & ...\\
\\
System D & \hi\  & 2.4737 & 13.50 $\pm$ 0.06 & 23.1 $\pm$ 3.1\\
& \ovi\ & 2.4738 & 13.53 $\pm$ 0.17 & 20.8 $\pm$ 9.8\\
& \civ\ & 2.4736 & 12.83 $\pm$ 0.05 & 50.4 $\pm$ 6.2\\
\\
System E & \hi\ & 2.4772 & 13.41 $\pm$ 0.05 & 22.1 $\pm$ 3.2\\
& \ovi\ & 2.4775 & 13.79 $\pm$ 0.11 & 23.8 $\pm$ 6.1\\
& \civ\ & ... & $\leq 12.13$ & ...\\
\\
System F & \hi\  & 2.5358 & 14.24 $\pm$ 0.89 & 23.1 $\pm$ 12.8\\
& \hi\  & 2.5365 & 15.26 $\pm$ 0.16 & 43.2 $\pm$ 7.0\\
& \ovi\ & 2.5358 & 13.04 $\pm$ 0.14 & 8.4  $\pm$ 3.7\\
& \ovi\ & 2.5365 & 13.96 $\pm$ 0.03 & 36.6 $\pm$ 3.3\\
& \civ\ & 2.5363 & $\le 12.28$  & ...\\
\\
System G & \hi\  & 2.5406 & 13.10 $\pm$ 0.05 & 17.8 $\pm$ 1.3\\
& \ovi\ & 2.5407 & 13.47 $\pm$ 0.07 & 12.2 $\pm$ 2.5\\
& \civ\ & 2.5406 & 12.42 $\pm$ 0.06 & 11.1  $\pm$ 2.3\\
\\
System H & \hi\  & 2.5422 & 13.31 $\pm$ 0.02 & 22.8 $\pm$ 1.1\\
& \ovi\ & 2.5422 & 13.62 $\pm$ 0.06 & 17.7 $\pm$ 2.7\\
& \civ\ & 2.5422 & 12.46 $\pm$ 0.07 & 16.6  $\pm$ 3.5\\
\\
System I & \hi\  & 2.5506 & 15.62 $\pm$ 0.91 & 25.1 $\pm$ 4.1\\
& \hi\  & 2.5511 & 15.64 $\pm$ 0.64 & 29.2 $\pm$ 3.6\\
& \ovi\ & 2.5506 & 13.37 $\pm$ 0.13 & 27.9 $\pm$ 9.8\\
& \ovi\ & 2.5510 & 12.97 $\pm$ 0.27 & 11.1 $\pm$ 6.2\\
& \civ\ & 2.5507 & 13.07 $\pm$ 0.03 & 8.9  $\pm$ 0.8\\
& \civ\ & 2.5510 & 13.52 $\pm$ 0.01 & 16.6 $\pm$ 0.6\\
\\
System J & \hi\  & 2.4571 & 13.07 $\pm$ 0.05 & 22.1 $\pm$ 1.6\\
& \hi\  & 2.4578 & 13.85 $\pm$ 0.01 & 33.8 $\pm$ 0.7\\
& \ovi\ & 2.4571 & 13.71 $\pm$ 0.16 & 5.7  $\pm$ 3.2\\
& \civ\ & 2.4571 & 12.09 $\pm$ 0.11 & 10.9 $\pm$ 3.8\\
\enddata

\label{colden}
\end{deluxetable}

\begin{deluxetable}{lccc}
\tablecaption{\hi-Only Systems in Proximity Regions }
\tablecolumns{5}
\tablewidth{0pt}
\tabletypesize{\scriptsize}
\tablehead{
\colhead{System} & \colhead{$z$} & \colhead{$\log$ \nhi(cm$^{-2}$)} & \colhead{$b$ (km s$^{-1}$)}
}
\startdata
{\it a} & 2.5193 & 13.65 $\pm$ 0.02 & 21.6 $\pm$ 0.9\\
{\it b} & 2.5240 & 13.71 $\pm$ 0.06 & 15.7 $\pm$ 1.4\\
{\it c} & 2.5244 & 13.74 $\pm$ 0.05 & 15.3 $\pm$ 1.5\\
{\it d} & 2.5393 & 14.42 $\pm$ 0.04 & 56.8 $\pm$ 2.8\\
{\it e} & 2.5433 & 14.44 $\pm$ 0.02 & 25.4 $\pm$ 0.4\\
{\it f} & 2.5485 & 13.63 $\pm$ 0.05 & 25.5 $\pm$ 1.7\\
{\it g} & 2.5493 & 14.30 $\pm$ 0.02 & 36.0 $\pm$ 1.7\\
{\it h} & 2.4556 & 13.56 $\pm$ 0.02 & 71.0 $\pm$ 2.3\\
{\it i} & 2.4591 & 13.76 $\pm$ 0.01 & 33.1 $\pm$ 0.5\\
{\it j} & 2.4634 & 14.07 $\pm$ 0.01 & 29.4 $\pm$ 0.6\\
{\it k} & 2.4722 & 14.44 $\pm$ 0.01 & 26.3 $\pm$ 0.2\\
{\it l} & 2.4737 & 14.29 $\pm$ 0.01 & 24.0 $\pm$ 0.3\\
{\it m} & 2.4746 & 14.21 $\pm$ 0.01 & 28.5 $\pm$ 0.4\\
\enddata

\label{coldenhi}
\end{deluxetable}

We used the software package
{\sc vpfit} \citep{carswell01} to determine redshifts, 
column densities $N$ and Doppler 
parameters $b$ (km~s$^{-1}$)  for a total of 23 absorption systems.
Table~\ref{colden} lists the results of the profile fits
for 10 systems with detected metal lines (labeled A--J),
and Table~\ref{coldenhi} for 13 H\,{\sc i}-only systems 
(labeled $a$--$m$). 
Portions of the spectra near selected transitions of interest, 
together with the {\sc vpfit} model fits, are reproduced in 
Figures~\ref{lines_2456} through \ref{lines_2457}. 
Again, we refer the reader to Figure~\ref{scheme}
for a schematic illustration of the location of the 
various absorption systems along the two lines of sight,
and to Figure~\ref{deltaz}
for an overall view of the \lya\ forest within the 
proximity regions.
We now briefly discuss each metal line system in turn.

\subsection{Systems A and B ($z=2.4554$ and $2.4573$)}

\begin{figure}
\plotone{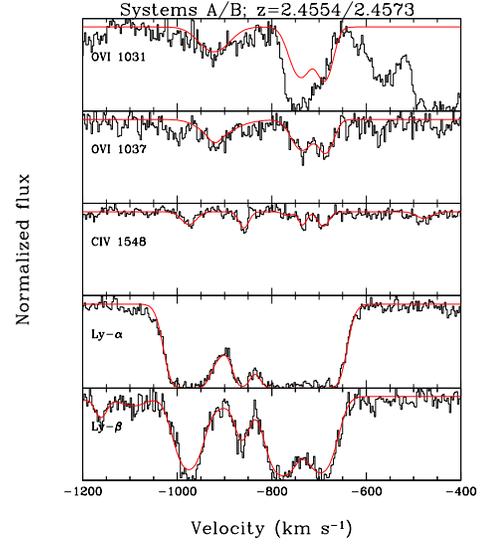}
\caption{Spectra (histograms) and fitted 
Voigt profiles (continuous red lines) of selected absorption
lines in systems A and B, as indicated. 
The $x$-axis velocity scale is relative to the systemic
redshift of the foreground QSO, as in Figure~\ref{deltaz}.
}
\label{lines_2456}
\end{figure}

\begin{figure}
\plotone{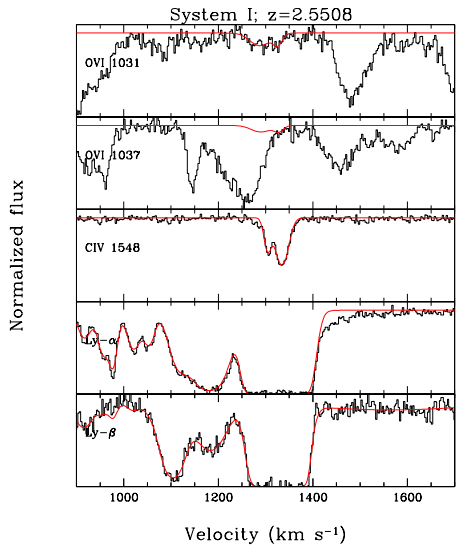}
\caption{Same as Figure \ref{lines_2456}, for System I. 
Unlike the previous cases, \civ\ absorption is very well defined
and stronger than \ovi, as may be expected for a system 
near the edge of the proximity region. \hi-only systems f and g appear near 1100 and 1200 \kms,
respectively (see Figure 3). }
\label{lines_2551}
\end{figure}

These two systems are very close together,
with physical separation of less than 1\,Mpc
(assuming negligible peculiar motions);
each of them consists of multiple absorption
components (see Figure~\ref{lines_2456}).
Associated \civ\ and \ovi\ absorption
is detected in both A and B. 
In the latter, there is a good match
in velocity between \civ\ and \ovi\
in the two components
at $z = 2.4578$ and 2.4583
(at $-735$ and $-692$\,km~s$^{-1}$ in
Figure~\ref{lines_2456}). 
At more negative velocities, however,
there is a single broad \ovi\ component
approximately mid-way between the \civ\
redshifts $z = 2.4550$ and 2.4564,
suggesting that this \ovi\ absorption has an origin
in diffuse, collisionally ionized gas, 
similar to the typical \ovi\ systems 
discussed by \citet{simcoe02}.

\subsection{System C ($z=2.4655$)}

\begin{figure}
\plotone{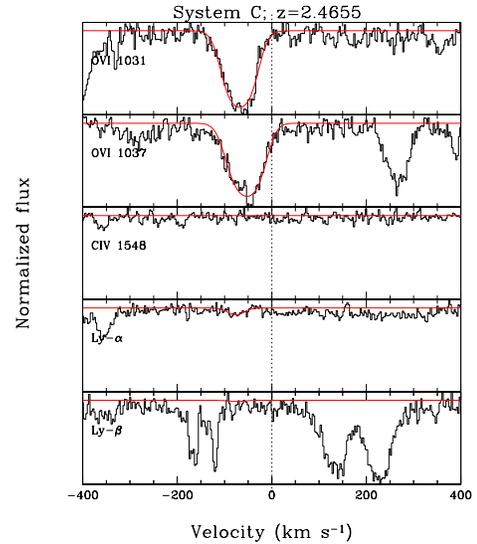}
\caption{Same as Figure \ref{lines_2456}, for system C. 
There is strong absorption in \ovi, and a weak \hi\ line. 
No other metal absorption lines are detected.}
\label{lines_2466}
\end{figure}

\begin{figure}
\plotone{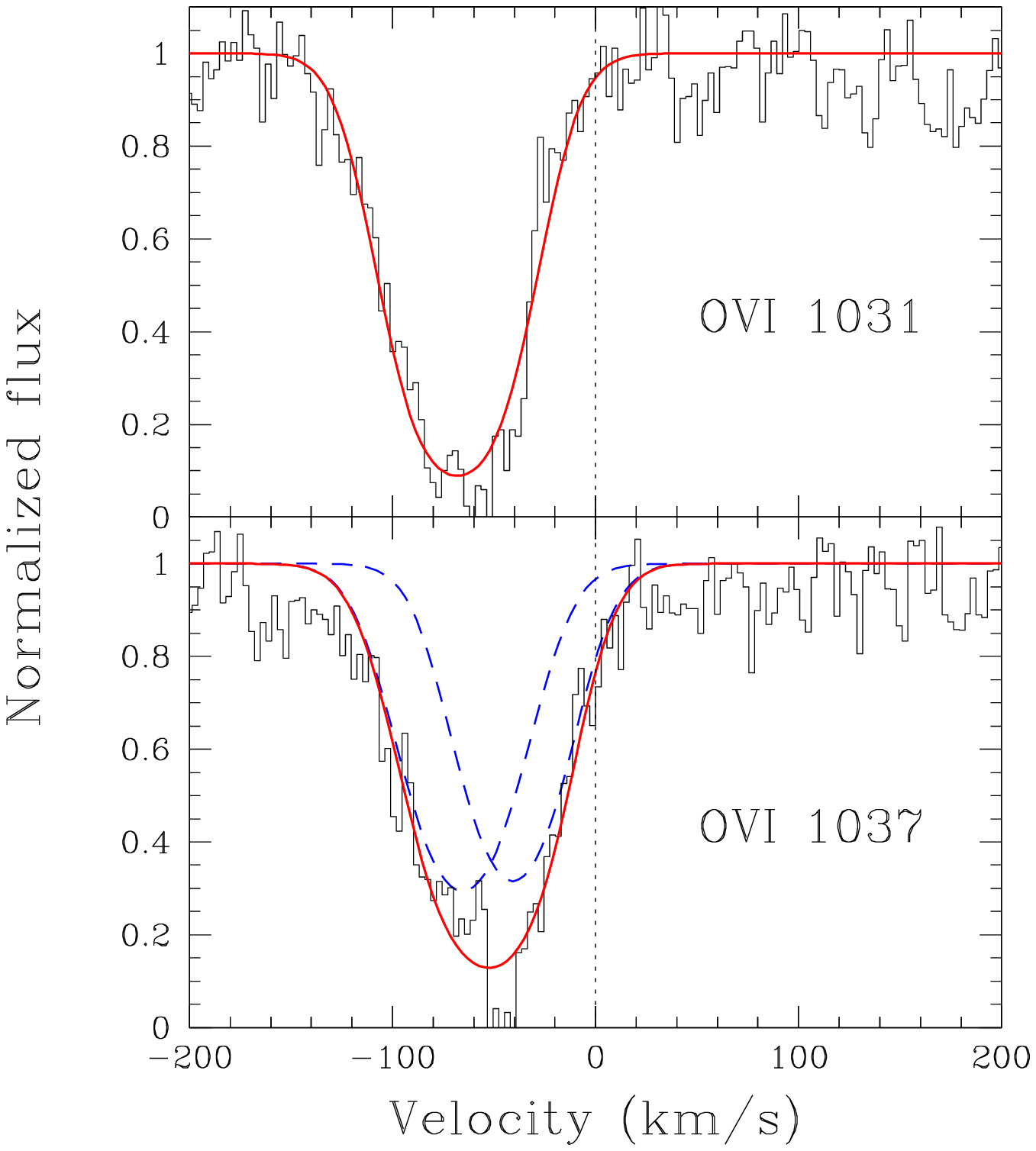}
\caption{\ovi\ absorption in system C (in the spectrum
of KP78 near the redshift of KP76).
As in the other figures, the histograms 
show the observed absorption profiles.
In the top panel, the red continuous line is the synthetic
absorption profile produced by VPFIT when fitting the 
\ovi\,$\lambda 1031$ transition alone. The corresponding
theoretical profile to the weaker member of the doublet
is shown in the lower panel (blue dashed line on the left). 
When combined with a lower redshift \lya\ line 
(blue dashed line on the right)
it produces a satisfactory fit to the blend (red continuous line).
The velocity scale on the $x$-axis is the same as in 
Figures~\ref{deltaz} and \ref{lines_2466}.
}
\label{kp76_OVI}
\end{figure}

This system is highly unusual in that it consists 
of a pair of very strong \ovi\ lines
with a very weak associated \lya\ line. 
No other metals lines are detected,
including N\,{\sc v} and \civ\ (see Figure~\ref{lines_2466}.) 
It is thus reasonable to question if our identification
is correct. One complication is that the weaker member
of the \ovi\ doublet, $\lambda 1037$, is partially blended
with a lower redshift \lya\ line (see Figure~\ref{kp76_OVI}). 
However, {\sc vpfit} returns the same values
of $z$, $\log N$(O\,{\sc vi}) and $b$
irrespectively of whether we fit the unblended 
O\,{\sc vi}\,$\lambda 1031$
line on its own, or simultaneously with the other member 
of the doublet and the blended \lya\ line.
In either case, VPFIT estimates the probability
that the fit represents the data to be greater than 90\%.
This absorption system is very close in velocity
to the systemic redshift of KP76
($\Delta v = 60$\,km~s$^{-1}$). 
Its unusual properties, with strong \ovi\ absorption, 
undetectable \civ, and barely detectable \lya\ 
(at the S/N of our data),
are consistent with the gas being exposed to a greatly enhanced
UV radiation field.
We return to this point in \S\ref{sec:plausibility}.

\subsection{Systems D and E ($z=2.4737$ and $2.4773$)}

\begin{figure}
\plotone{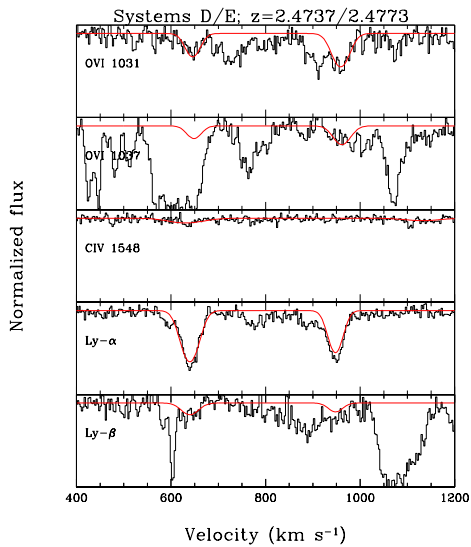}
\caption{Same as Figure~\ref{lines_2456}, for systems D and E. }
\label{lines_2474}
\end{figure}

These two systems present clear and isolated \lya\ absorption of moderate
strength (see Figure~\ref{lines_2474}). 
We identify associated \ovi\
absorption in both systems, although the 
\ovi\ lines are located in a relatively crowded region of the spectrum.
In particular, $\lambda 1037$ in system D is masked by a strong, 
saturated absorption feature (presumably \lya\ at a lower redshift).
\civ\ is barely detected in system D and below our detection limit in E.

\subsection{System F ($z=2.5362$)}

\begin{figure}
\plotone{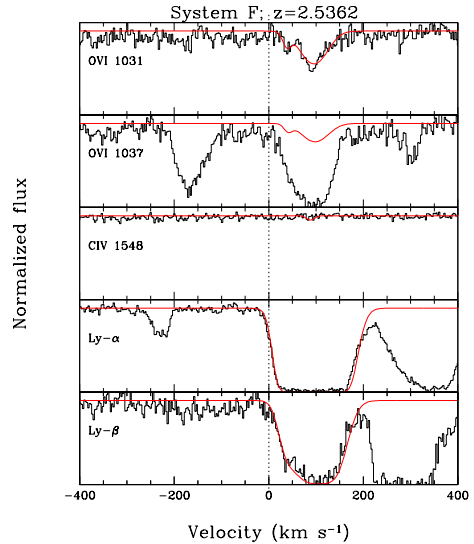}
\caption{Same as Figure \ref{lines_2456}, for System F. 
Hydrogen absorption is strong, as is O\,{\sc vi}\,$\lambda 1031$. 
O\,{\sc vi}\,$\lambda 1037$ is blended with hydrogen
lines at lower redshifts. 
C\,{\sc iv}\,$\lambda 1548$, on the other hand, is very weak
and below our detection limit.}
\label{lines_2536}
\end{figure}

\hi\ absorption is strong in this system, 
with $N$(H\,{\sc i})\,$=2 \times 10^{15}$\,cm$^{-2}$,
based on a strongly saturated \lya\ line and higher Lyman series
lines detected up to Ly$\delta$ (see Figure \ref{lines_2536}).
We clearly detect  O\,{\sc vi}\,$\lambda 1031$;
although $\lambda 1037$ falls within a stronger feature 
(presumably \lya\ at a lower redshift), 
the good redshift match of $\lambda 1031$ with the Lyman
lines lends support to our identification.
C\,{\sc iv}\,$\lambda 1548$ is very weak, below our
detection limit.
Again this system is very close in redshift to the foreground
QSO KP77 (see Figure~\ref{deltaz}); the velocity difference
is only $\Delta v = +76$\,km~s$^{-1}$.

\subsection{Systems G and H ($z=2.5406$ and $z=2.5422$)}

\begin{figure}
\plotone{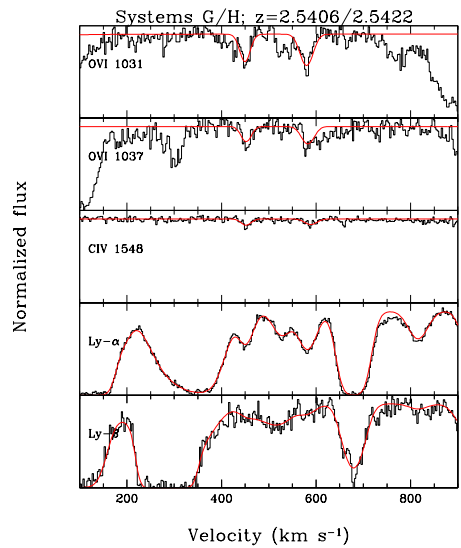}
\caption{Same as Figure \ref{lines_2456}, for Systems G \& H (\hi-only systems d and e are also
included in the plots, near 300 and 700 \kms, respectively, as well as several weaker Ly $\alpha$
components needed to fit the \hi\ profiles-- see Figure 3). 
\ovi\ lines are well defined and line up well with individual components within a complex
system of H\,{\sc i} absorption. As in
previous cases, C\,{\sc iv}$\,\lambda 1548$ is weak.}
\label{lines_2541}
\end{figure}

These two systems, separated by only $\sim 150$\,km~s$^{-1}$,
have very similar properties:
weak \lya\ and \civ, but relatively strong \ovi\
(see Figure~\ref{lines_2541}).
Such properties are unusual for most 
metal line absorption systems, but consistent
with a significant enhancement
in the local ionizing radiation field.

\subsection{System I ($z=2.5508$)}

This system is the furthest from the foreground QSO
among those considered here, with an inferred distance 
of $\sim 5$\,Mpc from KP77. 
Its properties are much more typical of intervening metal line systems,
with relatively strong C\,{\sc iv} and H\,{\sc i};
its column density
$N$(H\,{\sc i})\,$= 8.6 \times 10^{15}$\,cm$^{-2}$
is the largest among the absorbers
in the present sample.
Only the stronger member of the \ovi\ doublet is detected
($\lambda 1037$ is blended), but the proximity in velocity to
\civ\ and \hi\ supports the identification. 

\subsection{System J ($z=2.4575$)}

\begin{figure}
\plotone{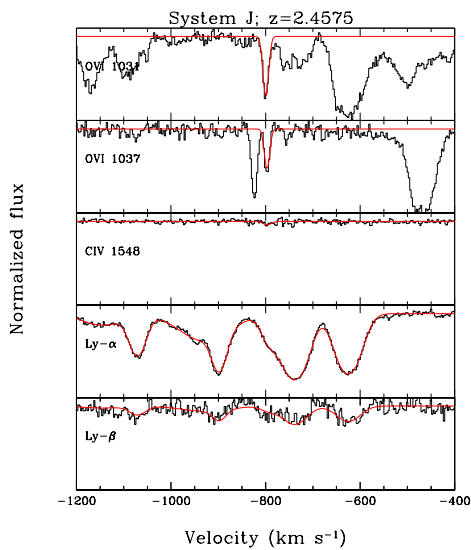}
\caption{Same as Figure \ref{lines_2456}, for System J. 
This system exhibits narrow \ovi\ absorption lines;
their width, when compared to that of the H\,{\sc i}
lines,  indicates that they are formed in gas at
$T \simeq 3 \times 10^4$\,K. The velocity range shown also includes \hi-only systems h and i (see 
Figure 3).}
\label{lines_2457}
\end{figure}

This is the only metal line system in the proximity region of 
of KP76 probed by the line of sight to KP77
(see Figure~\ref{deltaz}).
C\,{\sc iv}\,$\lambda 1548$ is marginally detected, 
while the O\,{\sc vi} doublet lines are strong and narrow
(see Figure~\ref{lines_2457}).
Interestingly, the $b$ values of the H\,{\sc i} ($z = 2.4571$ component), 
and O\,{\sc vi} lines (C\,{\sc iv} is too weak to provide
a reliable constraint on $b$) scale approximately in 
proportion to the square root of the atomic mass, 
as expected from thermal broadening alone (that is,
with no contribution to the line widths by large-scale
turbulence). 
The implied temperature is  $T \simeq 3 \ee{4}\units{K}$.

\section{THE EFFECTS OF THE QSO RADIATION FIELD}
\subsection{Inferences on the Radiation Field Intensity}

\begin{figure}
\plotone{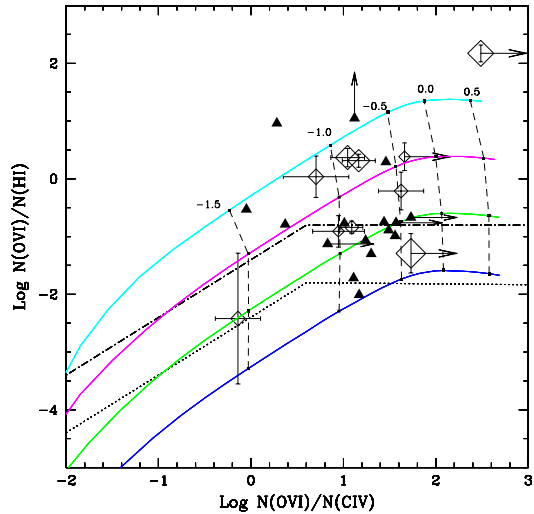}
\caption{Open symbols are \novi/\nhi\ vs. \novi/\nciv\ for 
metal line systems in the proximity regions of KP76 and KP77. 
Symbol size is proportional to the logarithm of the boost factor
$g$, listed in Table~\ref{gfactortab}.
Solid symbols are systems drawn from the \ovi-selected samples
of B02 and C02, for comparison (see text). Limits on ratios are
indicated with arrows. 
Also shown are the ion ratios calculated with the
photoionization code {\sc cloudy} for 
an assumed power-law radiation field  
$J(\nu) \propto \nu^{-\alpha}$, with $\alpha = 1.8$, 
a range of ionization parameters $U$, and four different values
of metallicity as follows: 
[O/H]\,$=-3$ (dark blue), [O/H]\,$=-2$ (green), [O/H]\,$=-1$ (magenta), 
and [O/H]\, $=0$ (cyan), with the usual convention
whereby [O/H]\,$= \log {\rm (O/H)} - \log {\rm (O/H)}_{\odot}$.
Vertical dashed lines join the loci of  
constant ionization parameter $U$, with values as
indicated. 
The dotted lines refer to H\,{\sc i}-only systems for which
we determine upper limits 
$\log N$(O\,{\sc vi})\,$\leq 12.7$ and
$\log N$(C\,{\sc iv})\,$\leq 12.1$; the dash-dotted line
is for systems with  $\log N$(H\,{\sc i})\,$ = 13.5$,
while the lower dotted line is for $\log N$(H\,{\sc i})\,$ = 14.5$.
}
\label{nratio_18}
\end{figure}

Most of the metal line systems
within the proximity regions of KP76 and KP77
exhibit strong O\,{\sc vi}, and relatively
weak C\,{\sc iv} and H\,{\sc i}.
In order to interpret these measurements
quantitatively and assess the impact of the 
ionizing flux from KP76 and KP77 on 
their environments, 
we used the photoionization code 
{\sc cloudy} \citep{ferland98} to model 
the ionization state of the gas in systems A--J.
The clouds were modeled as 
optically thin plane-parallel slabs, illuminated on one side
with a power-law radiation field ranging from 0.1 to 100 Ryd. 
Hydrogen number density 
and metallicity were allowed to vary. 
Figure~\ref{nratio_18} shows the measured 
column density ratios (or limits, where applicable)
together with the results of photoionization models
for four values of the metal abundance, from solar to 1/1000 of solar. 
The dashed lines show the line ratios expected
for a given value of the dimensionless 
ionization parameter $U$, defined as
\beq
U=\frac{\Phi}{n_{\rm H} \,  c}= \frac{n_{\gamma}}{n_{\rm H}},
\label{eq:U}
\eeq
where $\Phi$ is the flux of photons capable of ionizing H, $n_{\gamma}$ is
the ionizing photon number density, and $n_{\rm H}$ 
is hydrogen number density. 
We assume that the cloud is homogeneous so that the
column density ratios are independent of cloud size.

From Figure~\ref{nratio_18} it can be readily appreciated that,
for a given value of the ionization parameter, 
the ratio \novi/\nciv\ does not vary significantly 
with metallicity, 
provided of course that the O/C ratio remains constant---in
our modeling we have kept the relative abundance of the two elements
at the solar value. Consequently, the measured column density ratios
\novi/\nciv\ and \novi/\nhi\ can be used to constrain
the ionization parameter. Values of $\log U$ for each of the 
ten metal line systems are listed in column~8
of Table~\ref{gfactortab}, and can be seen to be mostly
between $\log U \simeq -1 $ and $0$.\footnote{
For simplicity, here and in the following discussion
we treat the upper limits to $N$(C\,{\sc iv}) as detections.
Thus the values of $U$ and $n_{\rm H}$ in 
Table~\ref{gfactortab} for systems C, E, and F are,
respectively, lower and upper limits to the true
values of the ionization parameter and hydrogen density.
}
The typical uncertainties in $\log U$ given the assumed radiation field
shape are $\sim 0.2-0.3$. 
The implied metallicities of the clouds, constrained by the ratios
\novi/\nhi\ and \nciv/\nhi, generally lie between a few tenths
of solar and a thousandth of solar---values that are 
typical of \lya\ forest clouds (e.g. Simcoe et al. 2004 [hereinafter S04]).
One exception is system C which exhibits an unusually high
column density of O\,{\sc vi} compared to H\,{\sc i}
and C\,{\sc iv}: its location on the diagram in Figure~\ref{nratio_18} 
cannot be reproduced even by the models with solar metallicity.
Also shown in Figure~\ref{nratio_18} are points drawn from a combination
of the samples of \citet{bergeron02} (B02) and \citet{carswell02} (C02), consisting
of systems selected by \ovi\ absorption only, without regard to \nhi.  
In general, nothing is known about the environment of the \ovi\ systems in the B02+C02 comparison
sample, although one of the systems from B02 is within 1500 \kms\ 
of the published QSO emission redshift and has only a lower limit on log \novi $/$\nciv\
implying that $\log U > -0.5$, similar to several of the proximate systems in our
sample.  We also note that there are two systems (out of 16) from the B02+C02 sample 
with implied metallicities similar to that
of system C, albeit with inferred values of $\log U$ that are lower by $\sim 2$ dex. 
We return to a comparison of the properties of the proximate systems versus published ``non-proximate'' samples
\footnote{We refer
to these samples as ``non-proximate'', meaning ``not known to be proximate''. It is possible that some
systems in the comparison samples are being ionized by more than the metagalactic radiation field.} 
below.

\begin{deluxetable*}{lccccccccc}[!tbp]
\tablecaption{Proximate Absorption System Model Results}
\tablecolumns{2}
\tablewidth{0pt}
\tabletypesize{\scriptsize}
\tablehead{
\colhead{System} & \colhead{$z$} & bQSO\tablenotemark{a} & fQSO\tablenotemark{b} & \colhead{$r_{\rm
Q}$}\tablenotemark{c} &
\colhead{$\Delta t$\tablenotemark{d}} &
\colhead{$g$\tablenotemark{e}} 
& \colhead{$\log U$\tablenotemark{f} } 
& \colhead{$\log n_{\rm H}$\tablenotemark{g}}
& \colhead {$\log \rho/\rho_0$\tablenotemark{h}}\\
& & & & \colhead{(Mpc)}  &  \colhead{(Myr)} & & & (cm$^{-3}$) & ($z=2.5$)  
}
\startdata
A        & 2.4554 & KP78 & KP76 & 3.83 & ~0.2 & ~~7.9 & $-$1.05 & $-$3.4 & 1.6\\
B        & 2.4573 & KP78 & KP76 & 3.28 & ~1.2 & ~12.4 & $-$0.90 & $-$3.4 & 1.6\\
C        & 2.4655 & KP78 & KP76 & 1.06 & ~2.6 & 104.1 & $+$0.55 & $-$3.9 & 1.1\\
D        & 2.4737 & KP78 & KP76 & 2.69 & 16.8 & ~16.1 & $-$1.20 & $-$3.0 & 2.0\\
E        & 2.4773 & KP78 & KP76 & 3.86 & 24.5 & ~~7.8 & $-$0.35 & $-$4.1 & 0.9\\
F        & 2.5362 & KP78 & KP77 & 1.38 & ~5.5 & 196.0 & $-$0.35 & $-$2.7 & 2.3\\
G        & 2.5406 & KP78 & KP77 & 2.20 & 12.9 & ~76.3 & $-$0.80 & $-$2.7 & 2.3\\
H        & 2.5422 & KP78 & KP77 & 2.64 & 16.0 & ~53.4 & $-$0.75 & $-$2.9 & 2.1\\
I        & 2.5508 & KP78 & KP77 & 5.23 & 33.4 & ~13.5 & $-$1.55 & $-$2.7 & 2.3\\
J        & 2.4575 & KP77 & KP76 & 3.19 & ~0.9 & ~11.5 & $-$0.50 & $-$3.8 & 1.1\\
{\it a}  & 2.5193 & KP78 & KP77 & 5.33 & ~0.5 & ~13.1 & ...& ...& ...\\
{\it b}  & 2.5240 & KP78 & KP77 & 3.87 & ~0.8 & ~24.8 & ...& ...& ...\\
{\it c}  & 2.5244 & KP78 & KP77 & 3.75 & ~0.8 & ~26.4 & ...& ...& ...\\
{\it d}  & 2.5393 & KP78 & KP77 & 1.89 & 10.5 & 104.2 & ...& ...& ...\\
{\it e}  & 2.5433 & KP78 & KP77 & 2.95 & 18.1 & ~42.7 & ...& ...& ...\\
{\it f}  & 2.5485 & KP78 & KP77 & 4.52 & 28.7 & ~18.2 & ...& ...& ...\\
{\it g}  & 2.5493 & KP78 & KP77 & 4.76 & 30.3 & ~16.3 & ...& ...& ...\\
{\it h}  & 2.4556 & KP77 & KP76 & 3.82 & ~0.8 & ~~8.0 & ...& ...& ...\\
{\it i}  & 2.4591 & KP77 & KP76 & 2.69 & ~1.0 & ~16.2 & ...& ...& ...\\
{\it j}  & 2.4634 & KP77 & KP76 & 1.49 & ~1.7 & ~52.9 & ...& ...& ...\\
{\it k}  & 2.4722 & KP77 & KP76 & 2.30 & 13.9 & ~22.2 & ...& ...& ...\\
{\it l}  & 2.4737 & KP77 & KP76 & 2.75 & 17.0 & ~15.5 & ...& ...& ...\\
{\it m}  & 2.4746 & KP77 & KP76 & 3.04 & 18.9 & ~12.7 & ...& ...& ...\\
\enddata
\tablenotetext{a}{Background QSO spectrum in which absorption is measured.}
\tablenotetext{b}{Foreground QSO within ``proximity region''.}
\tablenotetext{c}{Distance from fQSO.}
\tablenotetext{d}{Difference between the time corresponding to the observed QSO luminosity and
the time at which the physical conditions in the gas are being measured by the background QSO light (see
text for discussion).}
\tablenotetext{e}{Enhancement of the ionizing radiation field under the assumption that the fQSO
is radiating isotropically and had the same luminosity $\Delta$t Myr ago.}
\tablenotetext{f}{Inferred ionization parameter based on photoionization models.}
\tablenotetext{g}{Inferred hydrogen density, assuming that the radiation field
intensity is produced by the foreground QSO and that it is a factor $g$ larger than the metagalactic
background.}
\tablenotetext{h}{Hydrogen density in units of the mean density, assuming $\Omega_b=0.04$. }

\label{gfactortab}
\end{deluxetable*}

Having determined the value of $U$ for each metal line
system, we can now deduce the density
$n_{\rm H}$  corresponding to
a given value of $\Phi$ in eq.\,(\ref{eq:U}). 
In particular, we are interested in examining
the consequences of assuming that the 
ionizing flux seen by each cloud  
consists of the sum of two components: 
\textit{(i)} the metagalactic ionizing background and 
\textit{(ii)} the radiation field of the QSO in the proximity zone of which
the absorber is located. Specifically, we consider the scenario
whereby the metagalactic background to which the cloud
is exposed is boosted by the 
presence of the nearby QSO by a factor $g$, given by:
\beq
g=1+\frac{10^{-0.4(48.60+m_{912}) }}{(1+z)\pi J_{\rm bg}(\nu_0)}\left\lbrack\frac{d_{\rm L}(z)}{r_{\rm Q}}\right\rbrack^2,
\label{eq:gfactor}
\eeq
where $d_{\rm L}(z)$ is the luminosity distance to the QSO
from Earth, and $J_{\rm bg}(\nu_0)$ is the
background flux density at the Lyman limit; 
we adopt 
$J_{\rm bg}(\nu)\sim 5\times10^{-22}(h\nu/13.6\units{ev})^{-1.8}$
erg s$^{-1}$ cm$^{-2}$ sr$^{-1}$, which is consistent with the inferred hydrogen
photoionization rate required to explain the mean Lyman $\alpha$ forest optical depth
\citep{bolton05}, and is approximately equal to the measured sum of the QSO and LBG contributions
to the metagalactic radiation field at $z \sim 3$ \citep{shapley06,hunt04}. It is a factor of
$\sim 2$ lower than the value of $J_{\nu}$ inferred from the QSO proximity effect \citep{scott00}, but
is consistent within the uncertainties. Clearly, our results depend on the assumed intensity and spectral
shape of the metagalactic radiation field, since we are seeking evidence for departures from these values
produced by local sources of ionizing photons. The normalization is probably uncertain by at least a factor
of $\sim 2$, so that inferred values of $g$ should be viewed as similarly uncertain. 
The value of $r_{\rm Q}$ is the proper distance of the 
absorber from the nearby QSO, 
obtained by combining the transverse 
distance implied by their angular separation on the sky 
and the distance along the line of sight 
inferred from the redshift space velocity difference $\Delta v$: 
\begin{equation}
r_{\rm Q}  = \left [ R^2(z) + \left ({\Delta v \over H(z)} \right )^2 \right ]^{1/2}
\end{equation}
where $R(z)$ is the proper distance 
on the plane of the sky between the foreground QSO
and the sightline to the background QSO 
and $H(z)$ is the Hubble parameter; both quantities are
evaluated at the redshift of the foreground QSO.  
As usual, the second term is affected by uncertainties due
to departures from the Hubble flow caused by local density fields and 
by the systematic uncertainties
in the QSO redshifts; together they may amount to less
than $\sim 1$\,Mpc  ($\simeq 260$ \kms) \citep{adelberger04}.
Values of $r_{\rm Q}$ for all 23 absorption systems
(metal line systems and H\,{\sc i}-only
systems) are given in column 5 of Table~\ref{gfactortab};
given our choice of maximum value of $\Delta v$
(see \S3) and the QSO separations,
they range from $\sim 1$\,Mpc  to $\sim 5$\,Mpc.

It is important to recall at this point that,
as discussed in detail by \citet{adelberger04}, 
the geometry of the TPE is such that 
there is a time difference between, on the one hand, the time when 
we measure the physical conditions of an absorbing cloud 
in the spectrum of a background QSO and, on the other,
the time when we measure the flux of the foreground QSO.
The time difference $\Delta t$  depends on both
the redshift and the transverse separation. 
For a velocity difference $\Delta v$ between
the foreground QSO and an absorption system 
at proper separation $r_{\rm Q}$, 
\begin{equation}
\Delta  t =  {1 \over c} \left[ r_{\rm Q}  + {\Delta v \over H(z)} \right] .
\end{equation}
Note that this time difference is {\it not} the same 
as the light travel time between the foreground
QSO and the gas giving rise to the absorption,
unless $z_{\rm abs} = z_{\rm fQ}$ (where $z_{\rm fQ}$
is the redshift of the foreground QSO). 
The value of $\Delta t$ for 
absorption systems with $\Delta v \gg 0$ 
(i.e., systems redshifted relative to the foreground
QSO) will be nearly twice 
the light travel time $r_{\rm Q}/c$, 
whereas $\Delta t < r_{\rm Q}/c$ 
for systems with $\Delta v \ll 0$.  
In other words, absorption systems within the proximity region
of a foreground QSO, but located behind it (as viewed from Earth),
probe physical conditions at times up to $\sim 30$\,Myr earlier 
than the time we receive the QSO light on Earth. 
Conversely, systems in front of the 
QSO can provide a more recent  
measure at a given $r_{\rm Q}$.  
Values of $\Delta t$ for all 23 systems  
considered in the two proximity regions of the foreground
QSOs KP76 and KP77 are collected in column 6 of Table~\ref{gfactortab}; 
they range between 0.2\,Myr and $33.4$\,Myr. 

Returning to the QSO boost factors $g$ given
by equation~(\ref{eq:gfactor}), their values, 
listed in column 7  of Table~\ref{gfactortab},
were calculated under the assumption that 
the QSOs radiate isotropically and have 
maintained near-constant luminosity for the corresponding
time intervals $\Delta t$. 
The lifetimes of bright QSOs are notoriously uncertain, 
but, as mentioned above, lifetimes of 1-100 Myr are believed to
bracket the typical values.  
For the time being, we use our naive determinations of $g$ 
and examine the results for consistency with
expectations.  In particular, the calculated values of 
$g$ and $U$ lead, via equations~(\ref{eq:U}) and (\ref{eq:gfactor}),
to the volume densities $n_{\rm H}$ listed in the
penultimate column of Table~\ref{gfactortab};
the last column gives the corresponding overdensities
relative to the mean density of baryons at $z = 2.5$.
From the Table we see that the clouds probed in the
proximity regions of KP76 and KP77 trace
densities relative to the cosmic mean of 
$\log \rho/\rho_0 = 0.9$--2.3; 
these are typical values for 
metal line systems of intermediate \hi\ column density
in the general IGM (e.g, S04). 

\begin{deluxetable*}{lccc}
\tablewidth{0pt}
\tabletypesize{\scriptsize}
\tablecaption{Column Densities With Reduced Radiation Field Intensities\tablenotemark{a}}
\tablehead{
\colhead{System} & \colhead{\nhi}  & \colhead{\novi}  & \colhead{\nciv}
}
\startdata
A & 15.78 & 12.77 & 13.71\\
B & 16.39 & 13.09 & 14.13\\
C & 13.63 & 12.35 & 13.58\\
D & 14.81 & 11.40 & 13.28\\
E & 14.37 & 13.70 & 13.18\\
F & 20.08 & 10.10 & 13.01\\
G & 15.01 & 10.32 & 12.86\\
H & 15.11 & 11.16 & 13.20\\
I & 17.21 & 10.68 & 13.38\\
J & 15.11 & 13.22 & 13.32\\
\enddata
\tablenotetext{a}{Column densities of HI, \ovi, and \civ\ assuming the same physical
densities as in Table~\ref{gfactortab}, but with the ionization parameter 
reduced by the boost factor $g$
in each case.}

\label{colden2}
\end{deluxetable*}

\begin{figure}[!tbh]
\plotone{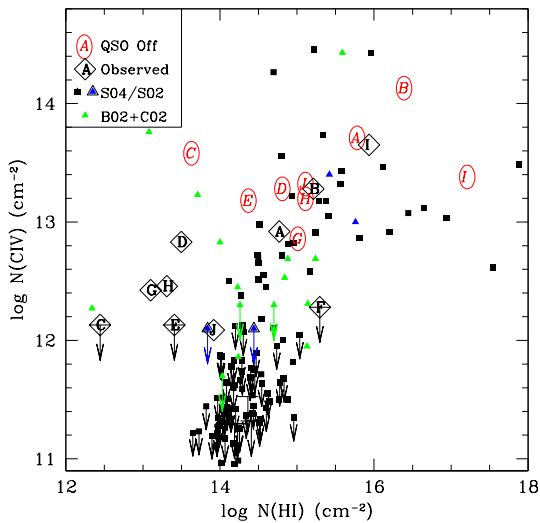}
\caption{Comparison between non-proximate 
absorption line systems from S04, B02,
and C02 and the proximate systems considered in this paper. 
The B02$+$C02 sample (green triangles), were
identified as \ovi\ systems without regard to their \nhi, whereas the 
S04 sample measured all systems with log \nhi $ > 13.6$. Upper limits on 
column densities are indicated with downward arrows.  
Diamonds labeled with the proximate system names 
are plotted at the measured values of column density,
while red ellipses are the values which we calculate
would obtain if the incident radiation field 
were decreased by the boost factors $g$ listed in
Table~\ref{gfactortab}. Blue triangles indicate 4 systems from S02 
that lie within 5000 \kms of the published QSO emission redshifts; the two
blue triangles with skeletal outline are systems for which we have accurate
QSO redshifts. The blue triangle near the observed system J corresponds to 
a proximate system near Q1700$+$64 with an estimated $g \simeq 120$ (see text).  
}
\label{figsimcoe_civ}
\end{figure}

\begin{figure}[[!tbh]
\plotone{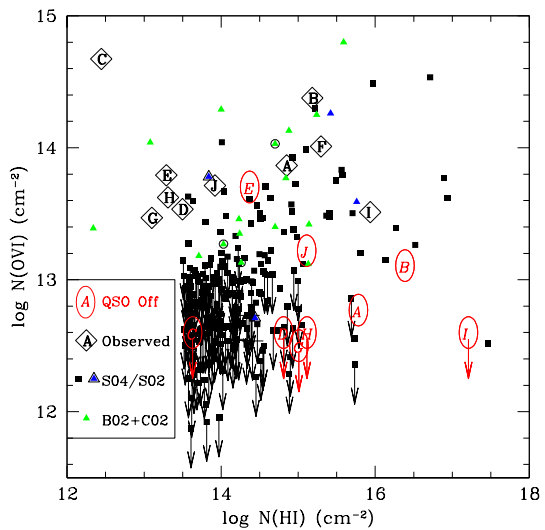}
\caption{As for Figure~\ref{figsimcoe_civ}, for \ovi. Red ellipses with downward arrows
indicate the approximate detection upper limit for \ovi\ in spectra similar to those used 
in this paper and in S04 (instead of their predicted \novi, which are much lower in some cases)
to facilitate comparison with the observed samples. As in figure~\ref{figsimcoe_civ}, the blue
triangle nearest to (observed) system J is a known proximate system near Q1700$+$64 with $g \simeq 120$.  }
\label{figsimcoe_ovi}
\end{figure}

With our photoionization modeling, we can further calculate
the column densities of H\,{\sc i}, C\,{\sc iv} and O\,{\sc vi}
which absorption systems A--J would produce, given 
their volume densities $n_{\rm H}$, in the absence of
the radiation field from the nearby QSO.
Using as input a radiation field now reduced by the
boost factors $g$ given by equation~(\ref{eq:gfactor}),
we obtained the column densities collected in  
Table~\ref{colden2}. 
In Figures~\ref{figsimcoe_civ} and \ref{figsimcoe_ovi}
we compare the ion column densities with and without the 
boost factors with the values measured
in \lya\ forest clouds by S04, supplemented by the smaller samples of B02
and C02 mentioned above.
Unlike the S04 sample, which measured or placed limits on \nhi\ and \novi\ for all systems with 
log \nhi $>13.6$, the latter studies searched for \ovi\ detections independently of
the associated \nhi, and thus may be more directly comparable to the sample of 10 proximate systems
in which \ovi\ is detected. 
Out of 16 \ovi\ systems in the combined B02+C02 sample, two (12.5\%) have log \nhi $<13.6$.
\citet{bergeron05} have presented a larger sample of 
\ovi-selected systems, based on 10 QSO sightlines with
column density sensitivity similar to B02+C02 and to the proximity regions in this paper.
With a total survey redshift path of $\Delta z \simeq 2$, these authors find  
that 7 of 51 (13.7\%) of the \ovi\ systems at $z = 2.3\pm0.3$ 
have log \nhi$<13.6$. Of the
51 \ovi\ systems, {\it all} have detected lines of \civ\ with log \nciv $\ge 12.30$ (systems
within 5000 \kms\ of the QSO redshifts were excluded from the sample). Note that, in contrast, 4 of the
10 proximate systems in our sample (C, E, F, and J) have log \nciv $<12.3$. 
Although the actual column densities
in the larger \citet{bergeron05} sample are not yet published, 
the overall distribution of \novi, \nciv, and \nhi\ appear 
similar to the samples included in Figures~\ref{figsimcoe_civ} and~\ref{figsimcoe_ovi}.   

Taking (for the moment) the observed properties of the proximate systems as
shown in Figures~\ref{figsimcoe_civ} and \ref{figsimcoe_ovi}, one sees that
many of the systems lie simultaneously at the lower end of the distribution of 
\nhi\ and the higher end of the distribution of \novi\ among all systems with 
detected \ovi. Only systems A, B, and I lie in regions inhabited by the main
locus of general intervening systems; system F, which is not anomalous in terms
of its \novi\ compared to \nhi, has significantly lower \nciv\ compared to any of the
S04 systems with comparable \nhi (figure~\ref{figsimcoe_civ}). 
Systems C, D, E, G, and H all have log \nhi $<13.6$, 
a trait we have seen is shared by only $\sim 13$\% of all systems with similar \novi. Based on the statistics
presented in \citet{bergeron05}, the expected number of such low-\nhi/strong-\ovi\ systems per unit redshift is
$dn/dz \simeq 3.5$, whereas in the proximate regions we find 5 such systems within a total 
redshift window of only $\Delta z \simeq 0.10$ (or $dn/dz \simeq 50$). 
Figures~\ref{figsimcoe_civ} and \ref{figsimcoe_ovi} also include 4 
\ovi\ systems which \citet{simcoe02} (S02) excluded from their analysis of 
strong \ovi\ systems because the absorption 
redshifts are within 5000 \kms\ of the QSO emission redshifts. In 2 of the 4 cases, we have accurate 
redshift and luminosity
information for the QSOs (Q1009$+$2956 and Q1700$+$6416), indicating that the two proximate
absorbers have $\Delta v=-2760$ \kms\ and
$\Delta v = -600$ \kms relative to the QSOs. Assuming negligible peculiar velocities, the corresponding
line--of--sight physical separations are $\sim 10.6$ Mpc and $\sim 2.2$ Mpc, with implied 
boost factors $g\sim 5$ and $g\sim 120$, respectively. These 2 systems were the only \ovi\ systems in the S02
sample which were {\it not} detected in \civ\ (see Figure~\ref{figsimcoe_civ}), and the latter system lies
very close to systems D, E, G, H, and (especially) J in both \nhi\ and \novi\ (figure ~\ref{figsimcoe_ovi}).   

If the radiation from
KP76 and KP77 were ``switched off'' (red symbols in Figures~\ref{figsimcoe_civ} and
\ref{figsimcoe_ovi})
\footnote{ System F, which is only 1.4\,Mpc from KP77,
has such a high boost factor ($g \simeq 200$) that
it would give rise to a DLA in the absence of
the radiation field from KP77; to improve
clarity we have not plotted this system for the
``QSO off'' case in Figures~\ref{figsimcoe_civ} and \ref{figsimcoe_ovi}.}, 
all of the proximate systems A-J are predicted to be detected in \civ\ and generally lie in the region
occupied by less-extreme intervening metal-line systems with intermediate \nhi. 
As seen in figure~\ref{figsimcoe_ovi}, 
only 4 of the 10 proximate systems (A, B, E, and J) are predicted
to have detectable \ovi\ in the ``QSO off'' case. These numbers are consistent with the \nhi-selected S04 sample--
for example, in the range $14.0 <$ log \nhi\ $ < 16.0$, the S04 sample has 116 systems (an average redshift
path density $dn/dz=56.6$ given the total path of the survey), of which 42\% are not detected in \ovi. 
In the total redshift path of the proximity regions considered here, $\Delta z \simeq 0.1$, one would
then expect $n \simeq 5.6$ systems in the same range of \nhi\ if it has an average density of
such systems (see \S 5 below). Six of the proximate systems are predicted to lie in this
range of \nhi\ for the ``QSO off'' case: A, D, E, G, J, and H, of which 3 (A, E, and J) are
predicted to yield \ovi\ detections.  
Similarly, the reduction of the assumed QSO-enhanced
radiation field intensities also lowers the number of predicted 
\ovi\  systems with log \nhi\ $<13.6$  within the proximity regions 
from 5 to 0, consistent with the expectation of only $n\simeq 0.35$ in $\Delta z\simeq 0.1$ from above.   

In terms of the predictions for individual systems, most move to less extreme
regions in both figure~\ref{figsimcoe_civ} and figure~\ref{figsimcoe_ovi} in the ``QSO off'' case (including
system F, which is not shown due to its large predicted \nhi). The exceptions are systems A and B,
which lie inside the locus of points from the literature for either case, and possibly system I.
The latter system, located at what we have considered
to be the ``edge'' of the QSO proximity region (at
a distance of 5.23\,Mpc from KP77), exhibits
$N$(C\,{\sc iv})/$N$(H\,{\sc i})
and $N$(O\,{\sc vi})/$N$(H\,{\sc i})
ratios which may be more in line with those
typically encountered in non-proximate systems
and actually appears slightly anomalous in both ratios
when the contribution of KP77 to the ionization rate
is removed (see Figures~\ref{figsimcoe_civ} and \ref{figsimcoe_ovi} ).
It may be significant that this system, at the largest $\Delta t$
among those considered here, reflects the QSO UV luminosity
33\,Myr prior to the time at which we observe KP77.
We return to this point in \S6.

In summary, none of the individual proximate metal-line systems is such an
extreme outlier (as compared to samples with similar \nhi\ or \novi\
from the literature) that explaining the observations {\it requires} an enhanced radiation
field from a nearby QSO. However, taken together, several statistical anomalies among the proximate sample
are eliminated when the systems are modeled using radiation field intensities
reduced by the $g$ factors in Table~\ref{gfactortab}.

\subsection{Plausibility of Enhanced Radiation Field Intensities}
\label{sec:plausibility}

To summarize the main conclusions of the analysis above,
we have shown with the aid of photoionization models that the
high degree of ionization exhibited by most absorption
system within the proximity regions of KP76 and KP77
can be explained in terms of an enhanced radiation field
and typical cloud densities (and metallicities).
The enhancements in the ionizing flux to which the clouds
are exposed are consistent with the values calculated from the
observed luminosities of the two QSOs at the Lyman limit
and the distances of the clouds from the nearby QSO,
ignoring the possibility that the QSO far-UV luminosities
may have varied significantly over the last $\sim 10^7$
years and that their emission may not be isotropic.
In this section we examine more carefully a number of factors 
which may affect this interpretation.

First of all, we consider whether it is plausible that the 
relatively high values of the ionization parameter indicated by the 
observed ion ratios may result from unusually low
densities $n_{\rm H}$ rather than abnormally 
high values of $n_{\gamma}$ (refer to equation~\ref{eq:U}). 
The densities so implied can be obtained straightforwardly by
dividing the values of $n_{\rm H}$ and $\rho/\rho_0$
in the last two columns of Table~\ref{gfactortab}
by the boost factors in column 7. In this scenario,
all of the clouds except one (system I) would be
at most mild overdensities relative to the cosmic
mean; in the most extreme cases (systems C, E, and F)
the absorption lines we see would arise in gas with density {\it lower}
than the cosmic mean (i.e., in voids). 
As pointed out by \citet{simcoe02},
such underdensities would imply proper sizes
in excess of 1\,Mpc; in such circumstances
it is difficult to understand how the O\,{\sc vi}
absorption lines would remain so narrow,
with $b$ values of 25--35\,km~s$^{-1}$, when
the differential Hubble flow across such large volumes 
would exceed $v_{\rm H} \simeq 250$\,km~s$^{-1}$.
We thus consider it more likely, given the presence
of nearby bright QSOs,  that many of the 
proximate systems are exposed to an enhanced flux
of ionizing photons, rather than arising in structures
of such low density.

One criticism which may be leveled at our photoionization
modeling is that it is rather simplistic in its underlying 
assumptions: we have assumed that H\,{\sc i}, C\,{\sc iv}
and O\,{\sc vi} absorption arises in the same gas at a 
uniform temperature and that the ion ratios are determined 
primarily by the balance between photoionization and 
recombination. In contrast, Simcoe et al. (2002) found that
the kinematics of strong \ovi\ absorption lines often differ 
from those of \civ\ lines at nearby velocities, and drew the
conclusion that much of the \ovi\ absorption
arises in collisionally ionized gas at a significantly
higher temperature than expected from photoionization
alone.
When we examine the proximate systems studied here,
however, we find that in most cases where C\,{\sc iv}
is detected the velocity match with O\,{\sc vi} is
very good, consistent with our assumption of photoionization
in the same parcels of gas. One exception was noted above
in system A (see Figure~\ref{lines_2456}), but in the other
systems the kinematics of the two ions are mutually
consistent.

\begin{figure}
\plotone{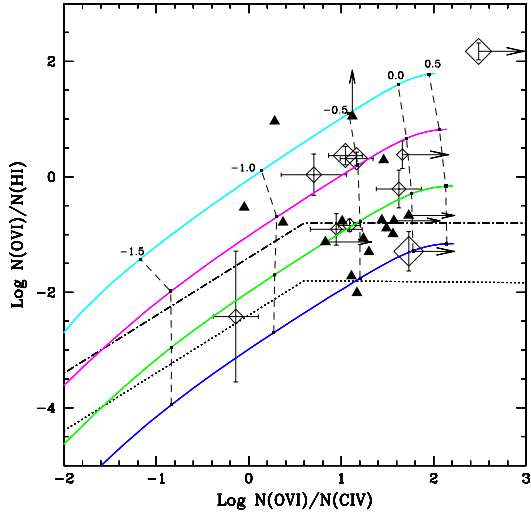}
\caption{Predictions of the {\sc cloudy} photoionization
models as in Figure~\ref{nratio_18}, but using a power law exponent
$\alpha=2.2$ for the incident radiation field.
In this case, the measured line ratios would imply
lower metallicities and higher ionization parameters
by factors of $\sim 2-3$. 
}
\label{nratio_22}
\end{figure}

Another simplistic feature of our modeling has been the
assumption that the spectral shape of the ionizing
radiation can be adequately represented by a power law
with spectral index $\alpha = 1.8$ 
[$J_{\rm bg}(\nu) \propto  \nu^{-\alpha}$; see \S4]
and that the nearby QSOs change only the intensity
and not the shape of the radiation field to which the
gas in each absorption system is exposed.
If the UV radiation field from a single QSO 
dominates over the metagalactic background, 
then variations in $\alpha$ with time, 
and between KP76 and KP77,
might result in different systems seeing
radiation fields of different shapes.   
Such variations may be the reason for the 
extreme ion ratios measured in system C
which would imply a super-solar metallicity
with the ionization spectrum we have assumed
(see Figure~\ref{nratio_18}).
For a given (O/H) ratio, the ratio
$N$(O\,{\sc vi})/$N$(H\,{\sc i}) 
depends strongly on the spectral shape of the
radiation field, as can be appreciated by comparing
Figures~\ref{nratio_18} and \ref{nratio_22}.
Changing the spectral slope $\alpha$ from 1.8 to
2.2 has the effect of lowering the metallicities
deduced from photoionization modeling 
by factors of $\sim 2-3$, because the fraction of
oxygen which is five times ionized increases.
On the other hand, this also has the effect
of increasing the ionization parameter $U$
required to reproduce the observed ion ratios,
thus implying lower values of $n_{\rm H}$
for a given boost factor $g$.

We have also investigated the effects on the ion ratios
of adopting the more realistic representation of the 
metagalactic background by 
Haardt \& Madau (private communication).
The radiation field calculated by these authors 
broadly resembles the $\alpha = 1.8$ 
power law over the wavelength range of interest, 
but also incorporates discrete spectral features 
resulting from radiative transfer effects in the IGM. 
With the Haardt \& Madau radiation field, we found that
\nhi\ remains approximately constant for a 
given value of $J_{\rm bg}(\nu)$, but that at a given
metallicity $\log N$(C\,{\sc iv}) and $\log N$(O\,{\sc vi})
are reduced by $\sim 0.2$ and $\sim 0.6$ respectively.
Thus, if the proximate systems considered here were
ionized by the metagalactic background rather than the
nearby QSOs, their high values of 
\novi/\nhi\ and \novi/\nciv\ would imply
even more unusual physical conditions (in terms 
of $n_{\rm H}$ and $\rho/\rho_0$) than 
deduced above under the power law assumption.  

Finally, we briefly discuss the 13 absorbers without
associated metal lines (systems $a$--$m$ in 
Tables~\ref{coldenhi} and \ref{gfactortab}).
The dotted and dash-dotted lines in Figures~\ref{nratio_18} and \ref{nratio_22}
show the loci corresponding to our detection limits
$\log N$(O\,{\sc vi})\,$ \leq 12.7$ and
$\log N$(C\,{\sc iv})\,$ \leq 12.1$ for two values of the 
hydrogen column density that bracket those measured
in systems $a$--$m$, $\log N$(H\,{\sc i})\,$ = 13.5$ (dash-dotted line)
and 14.5 (dotted line). The most straightforward interpretation
of the non-detections of 
O\,{\sc vi} and C\,{\sc iv} is that these
are low metallicity systems, with [O/H]\,$\simlt -2$.
On the other hand, without the diagnostics provided
by the O and C ion ratios, we cannot discern the 
effects of an enhanced radiation field on these
systems---it can be readily appreciated from 
Figures~\ref{figsimcoe_civ} and \ref{figsimcoe_ovi}
that the \hi-only systems are also entirely compatible 
with the values of $N$(C\,{\sc iv})/$N$(H\,{\sc i})
and $N$(O\,{\sc vi})/$N$(H\,{\sc i}) commonly
encountered in non-proximate systems 
with the same values of \nhi\ as we measure in the 
proximity regions. In the S04 sample, 62\% of the systems
with the same range in \nhi\ are not detected in \ovi\ to similar limits.

\section{THE LARGE SCALE ENVIRONMENT OF THE QSOs AND ITS IMPACT
ON THE PROXIMITY EFFECT}
\label{sec:environment}

\subsection{Galaxies and AGN in the Q1623+268 Field}
\label{subsec:gals}
\begin{figure}
\plotone{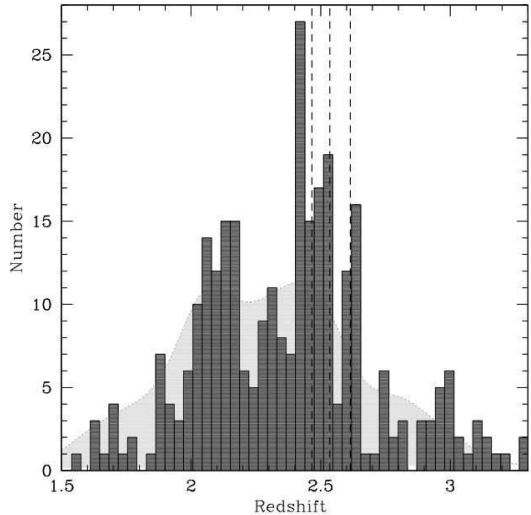}
\caption{
Redshift distribution of 298 
spectroscopically confirmed star-forming galaxies 
and AGN in the Q1623+268 field, selected by their
rest-frame UV colors. 
Galaxies have been grouped in redshift bins of
width $\Delta v =\pm 1500$\,km~s$^{-1}$,
so as to match the QSO proximity regions 
we have been considering. 
The redshifts of KP76, KP77, and KP78 are 
indicated with vertical dashed lines. 
The light, smooth curve in the background
shows the overall redshift selection function 
of a random sample of 298 objects selected using the
same color selection criteria (in the absence of clustering), 
based on the full spectroscopic sample of 
$\sim 2000$ galaxies in our on-going survey.
The three QSOs are located within moderate 
overdensities in the underlying distribution
of star-forming galaxies.
}
\label{nhist}
\end{figure}

One of the complications in the interpretation of the 
proximity effect, which has been appreciated since the
effect was first recognized, is the likelihood that QSOs
are preferentially found in overdensities in the matter
distribution. Any such density enhancement relative
to a more typical location in the IGM  could easily 
affect the statistics of \hi\ absorption on the physical
scales which are relevant to the proximity effect,
whether transverse or along the line of sight. 
In the present study, we have the advantage of knowing 
the large scale distribution of galaxies in the proximity
regions
of KP76 and KP77 from the spectroscopic
redshift survey that we have been
conducting in the Q1623+268 field---a full description
of the survey and its methods 
can be found in \cite{steidel04} and \cite{adelberger05}.
To date we have cataloged $\sim 300$
objects---star-forming galaxies and AGN---brighter
than ${\cal R} = 25.5$ at redshifts $z = 1.6$--3.3 over an area
of sky $\sim 11$\,\arcm\,$\times 15$\,\arcm\ approximately
centered on the KP76, KP77, and KP78 triplet.
The redshift distribution of the galaxies and AGN is
shown Figure~\ref{nhist}. The survey is far from
complete; in particular, in choosing objects for
spectroscopic follow-up we have given preference
to color-selected candidates at smaller projected 
distances from the QSOs on the plane of the sky.
Nevertheless, from the observed redshift
distribution we can 
estimate the overdensity
of galaxies and fainter AGN (in redshift space), 
compared to a sample drawn randomly
from the overall redshift selection function of the survey
(also shown in Figure~\ref{nhist}).

Within the field where spectroscopic follow-up has been
carried out, there are 19, 19, and 20 objects
with redshifts which place them within $\pm 1500$\,\kms\
of the systemic redshifts of, respectively, 
KP76, KP77, and KP78.\footnote{
Among these objects, three are faint
QSOs (${\cal R}=19.38$, 22.70 and 23.95)
at redshifts near that of KP76,
and one is a ${\cal R}=20.44$ QSO at a redshift 
near that of KP77.
This last one, Q1623-BX603, 
is located 97\arcs\ (780 kpc)
from the sightline to KP78.
With a redshift $z_{\rm em} \simeq 2.530$, 
estimated from its rest-frame UV emission lines, its
peak radiation would be expected near $-425$\,\kms 
(see middle panel of Figure~\ref{deltaz}) 
with a boost factor $g\simeq 7$ relative to the background.
If its redshift is correct (we have not obtained NIRSPEC 
spectroscopy of this QSO),
its contribution to the local radiation field at $-425$\,\kms\
is $\sim 10$\% of that of KP77. None of the other QSOs 
makes a significant contribution
to the radiation field in the proximity regions considered.
}
For comparison, the corresponding number of objects 
expected in the same redshift intervals for an unclustered 
population are 10, 9, and 6, respectively. 
Thus, the three QSOs appear to reside in moderate galaxy 
overdensities: ${\delta \rho / \rho} \simeq 1$ for the KP76 and KP77,
while KP78 lies in the most significant redshift space overdensity with 
${\delta \rho / \rho} \simeq 2$. 
However, these are the overdensities relative to the population
of star-forming galaxies at $z \sim 2.5$ (the BX galaxies
of Steidel et al. 2004) which themselves are 
significantly biased relative to the underlying mass distribution
at these redshifts:
\citet{adelberger05} measured a comoving 
correlation length $r_0 \simeq 4$ Mpc, which corresponds
roughly to a linear bias factor of $b \simeq 2$.
Thus, even regions containing an average density of BX galaxies
would probably represent overdensities in the matter distribution.
Of course, what we are interested in is the overdensity in 
H\,{\sc i} compared to an average location in the universe
at these redshifts. While it is uncertain how to relate
the overdensity of BX galaxies to that of H\,{\sc i}
[the factor relating the two presumably depends on
the threshold $N$(H\,{\sc i})], it seems reasonable
to conclude from the above that the Ly$\alpha$ forest
near KP76 and KP77 is denser by a factor of a few
compared to an average location in the IGM.

KP76, KP77, and KP78 are not unusual in being located
in galaxy overdensities; rather, their environments are
consistent with the galaxy-AGN cross-correlation function
measured from a much larger sample by \cite{as05}.
It is therefore not surprising that the small 
samples of QSO pairs used so far to search for the 
TPE have found little evidence to support it 
based on the expected decrease in H\,{\sc i} optical depth---it 
is easy to see how the overdensity of relatively high 
$N$(H\,{\sc i}) systems in the QSO environments
can more than compensate for the loss of 
absorption from optically thin systems 
whose \lya\ line equivalent
width would be most affected by an enhanced 
UV radiation field.  

\begin{figure*}
\plotone{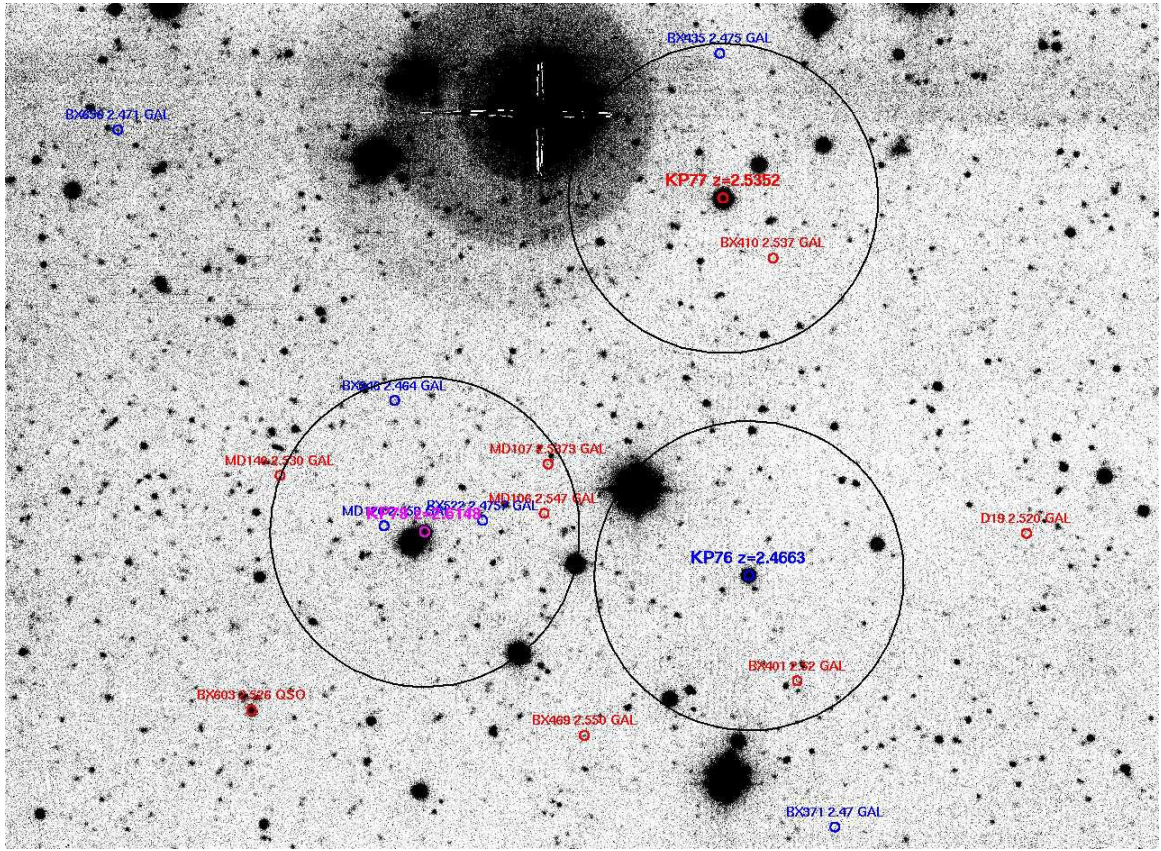}
\caption{$U_n$ band image of the 
Q1623+268 field showing the locations of 
the three bright QSOs and of spectroscopically
confirmed galaxies
within the $\pm 1500$\,km~s$^{-1}$ proximity regions 
of the two foreground QSOs. 
The black circles are 60\arcs\ in radius 
(corresponding to 0.48 proper Mpc in the 
transverse direction at $z = 2.5$). 
Galaxies and faint QSOs in the
proximity region of KP76 are shown in blue, 
while those in the proximity region of KP77 are labeled in red. 
North is up and East to the left.}
\label{field_image}
\end{figure*}

Strong correlations between galaxies and 
high \nhi\ absorption systems are known to exist on
scales of a few hundred  (proper) kpc at $z \simeq 2.5$ 
\citep{adelberger03,adelberger05,simcoe06}.
Figure~\ref{field_image} shows the locations of the three QSOs 
together with the objects from the spectroscopic
sample that lie within the proximity regions 
($\pm 1500$ \kms) of the foreground QSOs KP76 and KP77. 
Note that there are three galaxies within the 
proximity region of each of KP76 and KP77 
located within 60\,\arcs\ 
($\sim 0.5$ proper Mpc in the transverse direction
at $z = 2.5$) of the line of sight to KP78.
Similarly, there is one galaxy within the 
proximity region of KP76 located less than 0.5\,Mpc
from the sightline to KP77. 
The redshifts of the galaxies and their distances from the
QSO sightlines are listed in Table~\ref{galaxy_table};
the velocities of the galaxies relative to the foreground
QSOs are also indicated with light (yellow) shading
in Figure~\ref{deltaz}.
The galaxy redshifts were determined from
their rest-frame UV interstellar absorption lines
and/or \lya\ emission after correcting for the
systematic velocity offsets of these spectral features
from the systemic redshift---the error in this procedure
is $\pm 140$\,km~s$^{-1}$ (Steidel et al., in preparation).
For two of the galaxies, BX522 and MD107,  
we were able to measure the
systemic redshift directly from NIRSPEC 
observations of their H$\alpha$ emission lines
with a reduced error of $\pm 60$\,km~s$^{-1}$  \citep{erb06}.

\begin{deluxetable*}{lcccccc}
\tablewidth{0pt}
\tablecaption{Spectroscopically Identified Galaxies in fQSO Proximity Regions\tablenotemark{a}}
\tablehead{
\colhead{Galaxy} & \colhead{bQSO} & \colhead{fQSO}  & \colhead{$\Delta \theta$\tablenotemark{b}}  & \colhead{$d$ (kpc)\tablenotemark{c}}
& \colhead{$z_{\rm gal}$} & \colhead{$\Delta v_{\rm gal}$\tablenotemark{d}}
}
\startdata
MD126 & KP78  & KP76 & 16.6 & 133  & 2.458~ & $~-720\pm180$ \\ 
BX522 & KP78  & KP76 & 21.9 & 175  & 2.4757 & $~+810\pm~90$ \\
BX546 & KP78  & KP76 & 54.1 & 433  & 2.465~ & $~+200\pm180$ \\
MD106 & KP78  & KP77 & 46.2 & 370  & 2.547~ & $-1000\pm180$\\
MD107 & KP78  & KP77 & 54.1 & 433  & 2.5373 & $~+180\pm~90$ \\
MD140 & KP78  & KP77 & 60.2 & 482  & 2.530~ & $~-440\pm180$ \\
BX435 & KP77  & KP76 & 56.0 & 448  & 2.474~ & $+750\pm180$ \\
\enddata
\tablenotetext{a}{Galaxies within 60\arcs\ 
($\sim 0.5$\,Mpc) of background QSO sightlines and within the 
proximity region of the foreground QSOs---see Figure~\ref{field_image}. }
\tablenotetext{b}{Angular separation (in arcsec) between galaxy and 
background QSO (bQSO) sightline.}
\tablenotetext{c}{Projected physical distance between bQSO sightline and 
galaxy, at $z_{\rm gal}$.}
\tablenotetext{d}{Velocity difference (km~s$^{-1}$) 
between galaxy systemic redshift 
and that of the foreground QSO (cf. Figure~\ref{deltaz}). 
The quoted errors include estimates of the uncertainties 
of both the galaxy and the QSO systemic redshifts. }

\label{galaxy_table}
\end{deluxetable*}

Referring to Figures~\ref{deltaz} and \ref{field_image},
it is interesting to try and associate galaxies 
and absorption systems within the proximity regions
of KP76 and KP77.
The galaxy within a proximity region which lies
closest to the sightline to a background QSO is
MD126 (within the proximity region of KP76 and only
133\,kpc from the sightline to KP78).
Its redshift matches very well that of system B
(see Figure~\ref{deltaz}). As a matter of fact,
systems A and B exhibit similar ionization parameters
and metallicities (from our photoionization modeling
in Figure~\ref{nratio_18}), so that they could both
be due to MD126 (in which case their velocity difference
would be due to peculiar motions rather than the Hubble
flow as was assumed in calculating their distances
from the sightline to KP78).
The redshift of the next closest galaxy, 
BX522---175\,kpc from the KP78 sightline and also
in the KP76 proximity region, is intermediate 
between those of systems D and E 
which are separated by $\Delta v\simeq 300$\,km~s$^{-1}$
and both have metallicities [O/H]\,$\simgt -1$. 
The third galaxy within the proximity region of KP76,
BX546, is within 120\,\kms of system C, 
at a transverse distance of 438\,kpc.
Turning to galaxies close to the sightline to KP78 but
in the proximity region on KP77, 
MD 106 is within 130\,\kms of systems $b$ and $c$, 
while MD107 is within 100\,\kms of system F
(which in our modeling would have the properties of a DLA
in the absence of the radiation field from KP77).
Near the sightline to KP77, galaxy BX435 
is within 40 \kms of system {\it m}, 
and within the uncertainties 
has the same redshift as {\it  l} and {\it m}.
None of these possible associations between galaxies and absorption lines 
is particularly unusual or unexpected,  compared to regions lacking
bright QSOs but having similar galaxy overdensities 
\citep{adelberger03,adelberger05}. 
Similarly, given the level of incompleteness of the spectroscopic sample, 
the lack of identified galaxies associated with 
other absorption systems in the proximity regions 
should not be taken to imply that such galaxies are not present.  

\subsection{Matter Overdensities and the Proximity Effect}

\citet{faucher07} have recently presented a 
theoretical investigation of how the environments which
are likely to host QSOs may 
bias the inferred value of the hydrogen photoionization rate
from the metagalactic background, $\Gamma_{\rm bg}$, at
redshifts $z = 2$--4. 
These authors concluded that overdense environments
would lead to overestimates of $\Gamma_{\rm bg}$
by a factor of $\sim 2.5$ in determinations of the proximity
effect based on the distribution of 
Ly$\alpha$ optical depths $\tau_{\rm H\,I}$.
The bias arises from a combination of larger IGM density 
(resulting in larger values of $\tau_{\rm H\,I}$) and 
the infall of surrounding material onto the massive 
halos in which the QSOs are likely to reside. 
Such effects would lead one to conclude that the boost
by the QSO radiation to the ionizing
flux seen by nearby clouds is smaller than it really is, 
thereby leading to an overestimate of the metagalactic
background.

An additional complication is that 
essentially all previous measures of the proximity effect
have been based either on counting the number of
Ly$\alpha$ lines above a given equivalent width limit or
calculating the total transmitted flux 
near the redshift of the QSO, and comparing this statistic
with expectations for the general IGM at the same redshift.
This approach has been dictated by the fact that the
resolution of most spectra used to search for the proximity
effect is too coarse to determine $\tau_{\rm H\,I}$ directly.
The problem is that the response of the line equivalent
width to changes in $\tau_{\rm H\,I}$ is non-linear.
The lines which are most sensitive to changes in 
$\tau_{\rm H\,I}$ are those with $\tau_{\rm H\,I} < 1$
(on the linear part of the curve of growth), with equivalent
widths which are generally too small to be included
in the samples used to measure the proximity effect.
On the other hand, the Ly$\alpha$ lines which are included
typically have column densities 
in the range $14 \simlt \log N$(H\,{\sc i})\,$\simlt 17$,
where the equivalent width is a highly insensitive
measure of $\tau_{\rm H\,I}$ (such lines fall on the flat
part of the curve of growth).
Consequently,  the sensitivity of \lya\ line counts 
to a boost in the ionizing radiation field within the proximity regions
depends sensitively on line equivalent width. 
In spectra of moderate resolution and signal-to-noise ratio,
a situation may arise whereby the detection limit for
\lya\ line equivalent width is sufficiently high
that the increase in strong \lya\ lines due to the local
matter overdensity can mask an  
overall reduction in $\tau_{\rm H\,I}$ due to the enhanced
ionizing flux, except in very small regions very close to the QSO. 
In cases where the QSO redshift has been underestimated by
1000--2000\,km~s$^{-1}$ from rest-frame UV emission lines,
the region most affected by the QSO radiation field 
may not even be considered in 
typical proximity effect measurements!

In summary, the number per unit redshift
of the relatively strong
\lya\ lines that have been used in most previous 
measurements of the QSO proximity effect is likely to be 
closely related to the matter overdensity 
in the surrounding volumes.
While this gas will of course be affected by the QSO radiation field, 
the presence of the QSO could easily be
secondary to the local environment in 
dictating the statistics of such lines.
Once a more secure relationship is 
established between local galaxy density and the incidence
of relatively high $N$(H\,{\sc i}) systems, 
it may be possible to use the observed galaxy
density to calibrate out the environmental dependence 
on a QSO-by-QSO basis.  
On the other hand, because QSOs 
tend to inhabit relatively dense environments, one is more
likely to benefit from the presence of systems containing 
lines of highly ionized metals which, as shown here,
may offer the strongest observational constraints 
on the nature of the local radiation field.

\section{SUMMARY AND DISCUSSION}
\label{sec:discussion}

Using high resolution spectra of a triplet of QSOs 
with transverse separations
of $\sim 1$\,Mpc at $z \simeq 2.5$, 
together with accurate determinations of the QSO systemic
redshifts from rest-frame optical emission lines 
and extensive spectroscopic observations
of galaxies and AGN in the same field, 
we have conducted the most detailed investigation to date 
of the transverse proximity effect. 
By focusing on the regions of the IGM 
where the QSOs should overwhelm the metagalactic
radiation field by factors between  $\sim10$ and $\sim200$---if 
they have radiated isotropically and
with similar luminosities over the past 0.2--30\,Myr---we 
have examined the details of
the ionization state of individual metal absorption systems, 
rather than counting \lya\ lines above a given equivalent width
threshold, as has generally been done in previous attempts to measure 
the TPE.
We have shown that the 10 metal line systems
within the proximity regions of the two foreground QSOs have 
properties that are more easily explained if they are being illuminated 
by a UV radiation field significantly more
intense than the metagalactic background. 
Using photoionization models,
we have shown that most of the observed absorption systems 
have properties consistent with normal (i.e. non-proximate)
intermediate column density ($\log N$(H\,{\sc i})\,$\sim 14.5$--16.5) 
metal line systems that have been ionized by QSO continuum radiation 
with intensity consistent with
that inferred from the observed 
fluxes of the two foreground QSOs.   

We have placed the observed QSOs 
in the context of the large scale distribution of
galaxies in the same field, showing that all three lie in regions
of moderate galaxy overdensity---conditions typical
of high redshift QSOs and AGN. 
Even with accurate QSO redshifts (which correct the
published values for the same QSOs by 1000--2000\,km~s$^{-1}$), 
the naive expectation that 
there should be a dearth of \lya\ absorption systems 
in the proximity regions
of the foreground QSOs is not supported by the data. 
We argue that the environments of the
QSOs and an enhancement of moderate column density 
\hi\ absorption compared to
average locations in the IGM can easily mask the effects 
of the TPE if one relies on
counting statistics rather than examining 
the details of the gas-phase physical conditions.  

\subsection{QSO Lifetimes and Isotropy}
\label{sec:lifetimes}

In principle, it should be possible to use 
information on the distribution of the gas
that is clearly affected by QSO radiation 
relative to the positions and redshifts of the
foreground QSOs to measure, or set limits on, 
both the lifetime and the isotropy of
the QSOs radiation field.  
Each absorption system listed in Table~\ref{gfactortab}
and indicated in Figure~\ref{deltaz} samples a different time interval
$\Delta t$ and boost factor $g$ under
the simple hypothesis that the QSOs shine isotropically and
at constant luminosity over their radiative lifetimes. 
By comparing our sample to ``proximate'' and ``non-proximate'' absorption
systems from the literature, 
as in Figures~\ref{figsimcoe_civ} and  \ref{figsimcoe_ovi},
and from arguments based on cloud density and size,
we have shown that most of the systems in our sample---systems
C, D, E, F, G, H and J---are probably being over-ionized by 
the nearby QSO, while systems A and B are merely consistent with that hypothesis. 
Taken at face value, the spread of values of 
$\Delta t$  in Table~\ref{gfactortab} then implies
that the minimum QSO lifetime is
$\sim 25$\,Myr for KP76 (from system E) 
and $\sim 16$\,Myr for KP77 (from system H). 
As discussed above,
system I, in the proximity region for KP77, 
is the only system whose properties may favor 
ionization by the metagalactic field alone, without local enhancement; 
again, taken at face value,
this would indicate a radiative lifetime for KP77  in the range 
16\,Myr\,$<\Delta t < 33$\,Myr. 
Radiative lifetimes of 20--30\,Myr are entirely
consistent with He\,{\sc ii} TPE measurements 
by \cite{jakobsen03}, and with numerous estimates 
based on QSO duty cycle arguments 
(e.g., \citealt{steidel02}) or the local $M_{\rm bh} - \sigma$
relation between supermassive black holes and their 
host galaxies (e.g., \citealt{martini01}).  

In reality, there are many other variables which potentially
could modify such conclusions, even accepting 
the evidence that at least some of the absorption systems 
are over-ionized due to their proximity to a bright QSO.
For example, even if the lifetime of a QSO event is a 
well-defined quantity,  it is unlikely that
an accreting supermassive black hole maintains a constant luminosity 
throughout its active phase; rather, its accretion rate and UV output
could vary intermittently,  or grow (or decay) exponentially.  
Because each absorption system samples a different time in the QSO's history, 
it is entirely possible that systems whose distances from the QSO are larger 
(so that the $g$ factor is smaller) could have experienced a 
more intense radiation field in the past, or that a system at some
intermediate value of $\Delta t$ happened to coincide 
with a dormant period in the QSOs radiative history.  
We have also seen that the effects of an enhanced 
radiation field can be quite subtle. 
It is not always possible to say with confidence, even with high quality
data, whether or not a given absorber is experiencing
an enhanced ionizing radiation field, 
because we have no knowledge of its
environment or physical conditions prior to the time when the 
QSO began radiating at its present luminosity.

Given all these {\it caveats}, 
the strongest statement we can make about the isotropy
of the QSO radiation is that there is no evidence 
for anisotropy in the present data. 
If the QSOs' radiation were significantly beamed, 
one might reasonably expect to find absorption
systems well within the proximity zones 
with properties that are inconsistent with the assumption
of a radiation field intensity significantly
boosted over the metagalactic  background. 
Within the two proximity regions considered, 
the only absorption system 
that seems marginally inconsistent with the assumption of an enhanced
radiation field (system I) also has the 
largest value of $\Delta t$ and a 
relatively small boost factor $g$. 

As part of our survey of the Q1623+268 field, 
we obtained deep \textit{Spitzer} 
IRAC and MIPS images which include the 
QSO triplet.
We find that KP76, KP77 and KP78 have nearly identical
spectral energy distributions  
between 0.35 and $24\,\mu$m, 
with flat spectra ($f_\nu \simeq$\,constant) 
between 0.35\,$\mu$m and 4.5\,$\mu$m, and 
with $\nu f_{\nu}(0.36\mu{\rm m})/\nu f_{\nu}(24\mu{\rm m}) \simeq 4$. 
QSOs that are heavily obscured over a large solid angle would be
expected to be very bright in the thermal IR due to emission from heated
dust.
While modeling of the QSO spectral energy distributions
is beyond the scope of this paper, 
the data suggest that the two
foreground QSOs 
(as well as KP78, though it does not matter for the present purposes)
are not heavily obscured over a large fraction of 4$\pi$ steradians, given the
relatively weak 24$\mu$m luminosity. 
This provides independent support for
the hypothesis that these three QSOs would be seen 
as UV-bright over a large
fraction of a 4$\pi$ steradian solid angle.

The results we have reported provide a counter-example to recent 
claims of an excess of high \nhi\ absorption systems 
near the redshifts of foreground QSOs in the spectra
of background QSOs, 
compared to line-of-sight proximate 
absorbers with the same characteristics 
(e.g., \citealt{bowen06,hennawi06}). 
These authors interpret such an excess as evidence for
anisotropy of the QSO radiation which presumably (over)ionizes
the clouds in line-of-sight to Earth but is not seen 
(with the same intensity) by 
gas at transverse distances.
One difference that must be borne in mind is that
the analysis by Hennawi et al. (2006)
refers to absorption systems with column densities 
$\log N$(H\,{\sc i})\,$> 19$,  more than
3.5 orders of magnitude higher than any in the 
sample considered here. Similarly, the work
by Bowen et al. (2006) targeted strong Mg\,{\sc ii}
absorbers which are likely to be Lyman limit
systems with $\log N$(H\,{\sc i})\,$> 17.5$.
While our data appear to be inconsistent
with significant beaming of the ionizing
radiation from KP76 and KP77,
it is certainly possible that QSOs may differ
in the solid angle over which they radiate,
and that the apparent differences between our study
and those referenced above is due to such variations 
from QSO to QSO.
Finally, while we cannot rule out the possibility that there may be 
an anisotropic distribution of gas (as opposed to anisotropic
ionization) surrounding 
many QSOs, it would be hard to understand 
if such anisotropy extended over physical scales 
of $\sim 1$\,Mpc as would be required to explain some of the observations.

A lingering concern is the uncertainty in the
relevant distances introduced by the systematic
errors in the QSO systemic redshifts.
For example, it is debatable whether it is
surprising or not to find high $N$(H\,{\sc i})
absorbers near the redshift of a foreground QSO. 
Even in cases where the projected sightlines pass within
tens of kpc of one another, if the uncertainty in
the redshift of the foreground QSO
is $\sim 1500$\,km~s$^{-1}$  (equivalent to a 
distance of $\pm 6$\,Mpc at $z = 2.5$)
the relevant 
$g$ factor could be uncertain by up to three orders of magnitude.  
Many such ambiguities could be addressed 
with more accurate redshifts for
the QSOs, as well as more accurate values 
of $N$(H\,{\sc i}) and of other indicators 
of the physical conditions in
the gas.

\subsection{The Elusive Transverse Proximity Effect}

In this paper, we have used a particularly well-observed triplet 
of QSOs in a concerted effort
to test for the presence of the 
transverse proximity effect. We have shown that in spite
of the absence of an obvious ``clearing'' in the Ly$\alpha$ 
forest near the redshifts of bright
foreground QSOs, significant evidence for local 
enhancements in the ionizing radiation field
is present when one examines the detailed physical conditions
of metal line systems within 5\,Mpc of the foreground QSOs.
Crucial to this analysis are:  (a) the ability to measure
accurate H\,{\sc i} column densities 
(for which echelle spectra extending to at least Ly$\beta$ 
are required); 
(b) knowing as precisely as possible 
where to expect the influence of the foreground QSOs
(for which the forbidden line spectroscopy 
of the QSOs was essential);   (c) the ability
to detect weak lines of highly ionized metallic species, 
in this case \civ\ and \ovi; and (d) knowledge
of the large-scale environment inhabited 
by the foreground QSOs, since galaxy overdensities
are likely to be accompanied by 
the presence of appreciable \hi\ absorption.\footnote{Alternatively, as shown by
\citet{jakobsen03} and \citet{worseck06}, using the He\,{\sc ii} transitions
in concert with H\,{\sc i} is a very powerful technique 
because the effects of an enhancement in the
radiation field by a nearby QSO are then much more
evident than if one has access to H\,{\sc i} lines only.
However, the downside of this approach is that, given the 
short rest-frame wavelengths of the He\,{\sc ii} Lyman series,
there is only a handful of sightlines known at present
where this technique can be applied in practice---in most
cases intervening absorbers optically thick in the H\,{\sc i}
Lyman continuum prevent measurement of the He\,{\sc ii} lines. }

Essentially all previous searches for the TPE 
have been missing most or all of these ingredients.
As discussed above, reliance on relatively crude line 
counting or mean flux measurements 
in the forest, the likelihood that the redshifts of the foreground
QSOs, even when taken from carefully compiled 
catalogs such as SDSS, are incorrect by as much as 
2000\,km~s$^{-1}$ 
(an error $\delta z > 1000$\,km~s$^{-1}$ seems quite typical), 
and density enhancements of galaxies and intergalactic gas 
local to the foreground QSOs, 
could together conspire to mask the TPE.
In view of our results, these effects seem at least as 
plausible as anisotropic emission, short-timescale variability,
or very short QSO lifetimes 
in explaining the difficulties experienced so far in detecting the TPE. 
Based on the case investigated in this paper, 
the null hypothesis 
that bright QSOs radiate isotropically
over characteristic timescales of a few $10^7$ yr 
(timescales suggested by many, less direct, 
arguments) is consistent with the observations. 
Whether this is the rule rather than the exception 
could be established using similar observations of 
other multiple-QSO-sightline fields. 
Somewhat farther in
the future, the technique can be improved 
and extended using background \textit{galaxies} 
which would provide
much finer spatial and temporal sampling 
of the response of the IGM to 
radiation from QSOs \citep{adelberger04}.

\acknowledgements
We would like to thank 
Dawn Erb, Naveen Reddy, and Alice Shapley for their collaboration 
in the large survey which supplied the LRIS spectroscopic redshifts
of galaxies and AGN in the field of Q1623+268.
In addition, Dawn and Alice
obtained the near-IR spectra used to determine 
the systemic redshifts of the QSOs, 
and Naveen kindly provided the Spitzer results 
on the QSOs prior to publication.  
Kurt Adelberger is thanked for many discussions 
and for his participation in the early stages of the project. 
We would also like to thank 
Tom Barlow, Rob Simcoe, George Becker, and Bob Carswell for assistance 
with the software used to reduce HIRES data. 
Rob Simcoe also provided us with measurements
from his previous work, for which we are grateful. 
Wal Sargent provided us with his HIRES
spectra of KP76 and KP77 
which were combined with the new data presented here.
An anonymous referee, and Juna Kollmeier, provided constructive comments which significantly 
improved the paper. 
Finally, we wish to extend thanks to those of Hawaiian ancestry on whose sacred mountain
we are privileged to be guests. 
This work was supported by grants AST-0307263 and
AST-0606912 from the US National Science Foundation, 
and by the David and Lucile Packard Foundation.

\end{document}